\documentclass[a4paper,11pt]{article}
\pdfoutput=1 

\usepackage{jinstpub} 

\usepackage{amssymb}
\usepackage{amsmath}
\usepackage{wasysym}
\usepackage{array}

\usepackage{lineno}

\usepackage[figuresright]{rotating}

\usepackage{color}

\usepackage{subfigure}

\usepackage{hhline}

\def\co{\ensuremath{\textrm{CO}_2}}
\def\mcmaster{McMaster-Carr}
\def\labview{LabVIEW}
\def\np{pentane}
\def\Np{Pentane}
\def\tgc{sTGC}

\newcommand{\rr}{\raggedright}
\newcommand{\tn}{\tabularnewline}
\def\vp{vol\%}
\def\ml{mL}
\def\deg{\ensuremath{^\circ}C}

\title{Development and Characterisation of a Gas System and its
  Associated Slow-Control System for an ATLAS Small-Strip Thin Gap
  Chamber Testing Facility}

\author{R.~Keyes,}
\author{K.~A.~Johnson,}
\author{L.~Pepin,}
\author{F.~L\'eger,}
\author{C.~Qin,}
\author{S.~Webster,}
\author{A.~Robichaud-V\'eronneau,}
\author{C.~B\'elanger-Champagne,}
\author{B.~Lefebvre,}
\author{S.~H.~Robertson,}
\author{A.~Warburton\footnote{Corresponding author.},}
\author{B.~Vachon,}
\author{F.~Corriveau}
\affiliation{Department of Physics, McGill University, 3600 University Street, Montr\'eal, Qu\'ebec, Canada, H3A 2T8}
\emailAdd{awarburt@physics.mcgill.ca}

\abstract{
A quality assurance and
performance qualification laboratory was built at McGill University
for the Canadian-made small-strip Thin Gap Chamber (sTGC) muon detectors produced for the 2019-2020 ATLAS experiment muon spectrometer upgrade.
The facility uses cosmic rays as a muon source to ionise the quenching gas mixture of \np{} and
\co{} flowing through the \tgc{} detector. A gas system was developed
and characterised for this purpose, with a simple and efficient 
gas condenser design utilizing a Peltier thermoelectric cooler (TEC).
The gas system was tested to provide the desired 45~\vp{} \np{} concentration.
For continuous operations, a state-machine system was implemented with alerting and remote monitoring features to run all
cosmic-ray data-acquisition associated slow-control systems, such as high/low voltage, gas system and
environmental monitoring, in a safe and continuous mode, even in the absence of an operator.

}

\begin{document}
\maketitle
\flushbottom

\section{The Canadian \tgc{} Testing Facility for ATLAS}\label{lab}

The ATLAS~\cite{Aad:2008zzm} detector at the LHC~\cite{LHC} is scheduled for an upgrade of a subset of its muon detectors in
the 2019-2020 period~\cite{Kawamoto:1552862}. For this upgrade, small-strip Thin Gap Chambers
(\tgc{}s)~\cite{Kawamoto:1552862} produced in Canada will be tested
for quality assurance. A testing facility has been constructed at McGill University (Montr\'eal, Canada) for this purpose with
stringent requirements on performance and safety. 
The sTGCs designed for the ATLAS detector upgrade are planar multi-wire ionisation chambers operated in quasi-saturated mode, 
with an anode-to-cathode distance (1.4~mm) that is smaller than the distance between anode wires (1.8~mm). These chambers are operated 
with a \co{}:\np{} gas mixture (55\%:45\% by volume) at a typical high voltage (HV) of 3 kV.
This quenching-gas mixture makes it possible to operate the chamber in a high amplification mode by preventing the occurrence of streamers~\cite{Majewski:1984ag}.
An \tgc{} quadruplet module is made of a stack of four independent \tgc{}s in a single frame. Quadruplet \tgc{} modules designed for the ATLAS upgrade have an approximate 
size of 1 m (length) $\times$ 2 m (width) $\times$ 0.05 m (thickness) and a mass of about 70~kg.
While each \tgc{} is supplied with an independent HV connection, the four gas volumes of an \tgc{} quadruplet module are linked together during normal operation.

The Canadian \tgc{} testing laboratory is anticipated to characterise the performance of approximately 60 \tgc{} quadruplet modules over a period of 18 months.
To achieve this, a large cosmic-muon hodoscope has been installed at McGill University. 
The hodoscope structure measures 2.6~m $\times$ 2.6~m $\times$ 2.2~m (height) 
and can simultaneously hold 
in horizontal position up to four \tgc{} quadruplet modules on separate drawers stacked vertically.  
The top and bottom areas of the structure are covered by 2.5~cm thick scintillator sheets 
read out by photomultiplier tubes (PMTs). Signals from the scintillator detectors provide the trigger signal for the sTGC quadruplet modules readout.

The testing facility has a multi-channel HV system 
for the simultaneous operation of \tgc{}s, a multi-channel low voltage (LV)
system for the \tgc{} readout electronics, a custom trigger and data
acquisition system, and a gas system that can separately provide \co{} gas or a gaseous \np{} and
\co{} mixture. Pure \co{} gas is needed to purge detectors of their oxygen,
which should not be mixed with \np{} for safety reasons, and other possible contaminants.
It is also used to remove the \np{}
from the detector once operations are terminated, to avoid potential damage to the \tgc{}s
that can occur if trace amounts liquefy inside the gas volume.
All these systems are monitored
and controlled by a slow-control system that additionally
monitors the environmental conditions of the testing facility.
A state machine~\cite{Bitter} in the slow-control system has been developed to ensure a
safe operation of the testing facility.

The complete testing and performance characterisation of an \tgc{}
quadruplet module spans a period of a few to several days, during
which the gas system is required to operate continuously in a safe and
stable manner. More than one \tgc{} quadruplet module is expected to be
tested simultaneously, requiring the availability of several different
gas lines with the ability to deliver and control either \co{} or a \co{}:\np{}
gas mixture at different points in time.   
The design of the system
must allow the flexibility to modify the flow and type of gas provided
on individual gas lines, as well as the addition or removal of a gas connection, without having to stop the system.
The gas system is also
required to operate more or less continuously for a period of
approximately 18 months without any major downtime.

The paper is arranged as follows. Section~2
describes in detail the design, implementation and performance
of the gas system while Section~3 describes the design,
implementation and performance of the slow-control system of the
Canadian sTGC testing facility. A summary is provided in Section~4.

\section{Gas System}
The design of the gas system for the Canadian \tgc{} testing facility is
based on similar existing systems located in Israel (Tel Aviv
University, Weizmann Institute)~\cite{Etzion:2003hx}, Switzerland (CERN)~\cite{safety} and Canada (TRIUMF)~\cite{openshaw}. Unique constraints imposed by
the McGill laboratory space, such as physical size, building safety protocols, and laboratory amenities, require 
that specific procedures and functionality be incorporated into the system.

The gas system must provide a continuous flow with a stable concentration (within $\pm$3\vp{}) of either pure gaseous \co{} or gaseous \co{}:\np{}
mixture (55\%:45\% by volume, respectively) at input \co{} flow rates ranging from 50 to 100~\ml{}/min per line in up to 10 lines simultaneously. 
The system was designed
to have five lines providing independent flow rates of pure \co{} gas,
and five dual-purpose lines that can be operated independently with
either pure \co{} gas or the \co{}:\np{} mixture required for \tgc{}
operation. 
To avoid out-of-plane deformations and damage
to the \tgc{}s, the delivered pressure on each line should not exceed 0.5~kPa above the ambient pressure.
The \np{} exiting the chamber must be disposed of in a manner that is safe for the
laboratory and the surrounding environment. 
The system must be simple to operate and be equipped with automated monitoring and controls to permit running during extended periods when an operator may not always be present. The system must also be intrinsically fail-safe, which we define as a capacity to react to particular potential failure conditions by defaulting to a safe state that requires manual intervention before normal operations can resume.

Operational safety is central to the design of the gas system. Because 
there is no fume hood in the laboratory, no handling of liquid or gaseous \np{} is permitted to take place there, due to its
volatility and flammability. The laboratory is equipped with a
negative-pressure exhaust that is used to dispose of unreclaimed gas
downstream of the detectors. Within the gas system, all wetted parts must be constructed from materials resistant to
\np{}, which is known to attack rubber and
various plastics and coatings~\cite{Cole}. Otherwise, chemical residues could end up contaminating the \co{}:\np{} gas mixture and, in turn, the \tgc{}s.

\subsection{Design and Implementation}

The main component of the gas system is the mixing apparatus. It takes
pure \co{} gas and liquid \np{} as inputs and provides the desired
mixture as a gaseous output. Different methods can be considered to control
the output gas composition of a mixer, mainly by setting either the mixing volumes
or the mixture temperature. From maintenance and simplicity considerations, 
temperature control is chosen for this gas system, following the gas system designs
of the aforementioned facilities. 
Using the temperature dependence of the vapour pressure of liquid
\np{}, Amagat's and Dalton's laws for ideal gases~\cite{ThermodynamicsTextBook} (see Equation~\ref{eqn:Amagats}), which state that in a gaseous
mixture the volume ($V$) fraction of constituent gases is proportional to
their respective pressure ($p$) fractions, the volume fraction can in turn
be controlled for constituent gas $x$:

\begin{equation}
  \frac{V_{\rm x}}{V_{\rm tot}} =  \frac{p_{\rm x}}{p_{\rm tot}}.
  \label{eqn:Amagats}
\end{equation}

\begin{figure}
  \center
  \includegraphics[width=10cm]{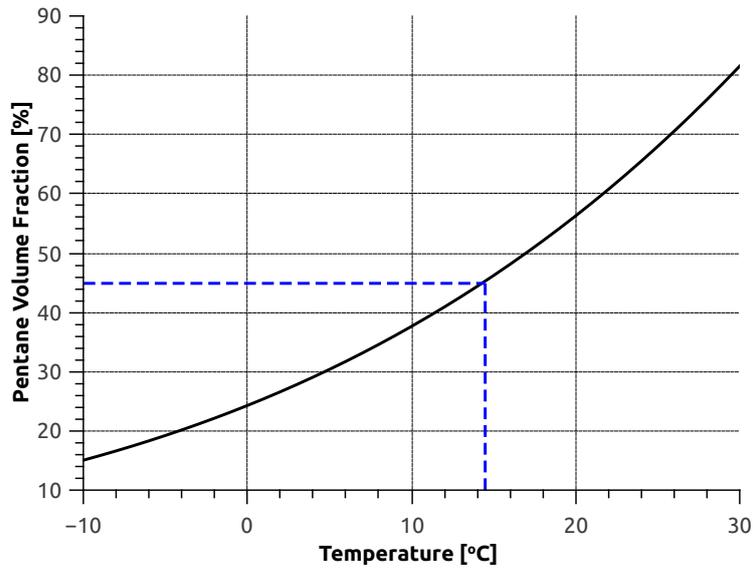}
  \caption{Volume fraction of \np{} as a function of temperature assuming atmospheric pressure and Amagat's law~\cite{CHERIC}. The blue dashed line indicates the desired 45~\vp{} \np{} operating point.}
  \label{fig:PentaneFraction}
\end{figure}

Assuming atmospheric pressure (101.325~kPa), Equation~\ref{eqn:Amagats} 
and the vapour pressure of \np{} as a function of temperature obtained from CHERIC~\cite{CHERIC}, 
Figure~\ref{fig:PentaneFraction} displays the temperature dependence of the \np{} volume fraction.

\begin{figure}
  \center
  \includegraphics[width=12cm]{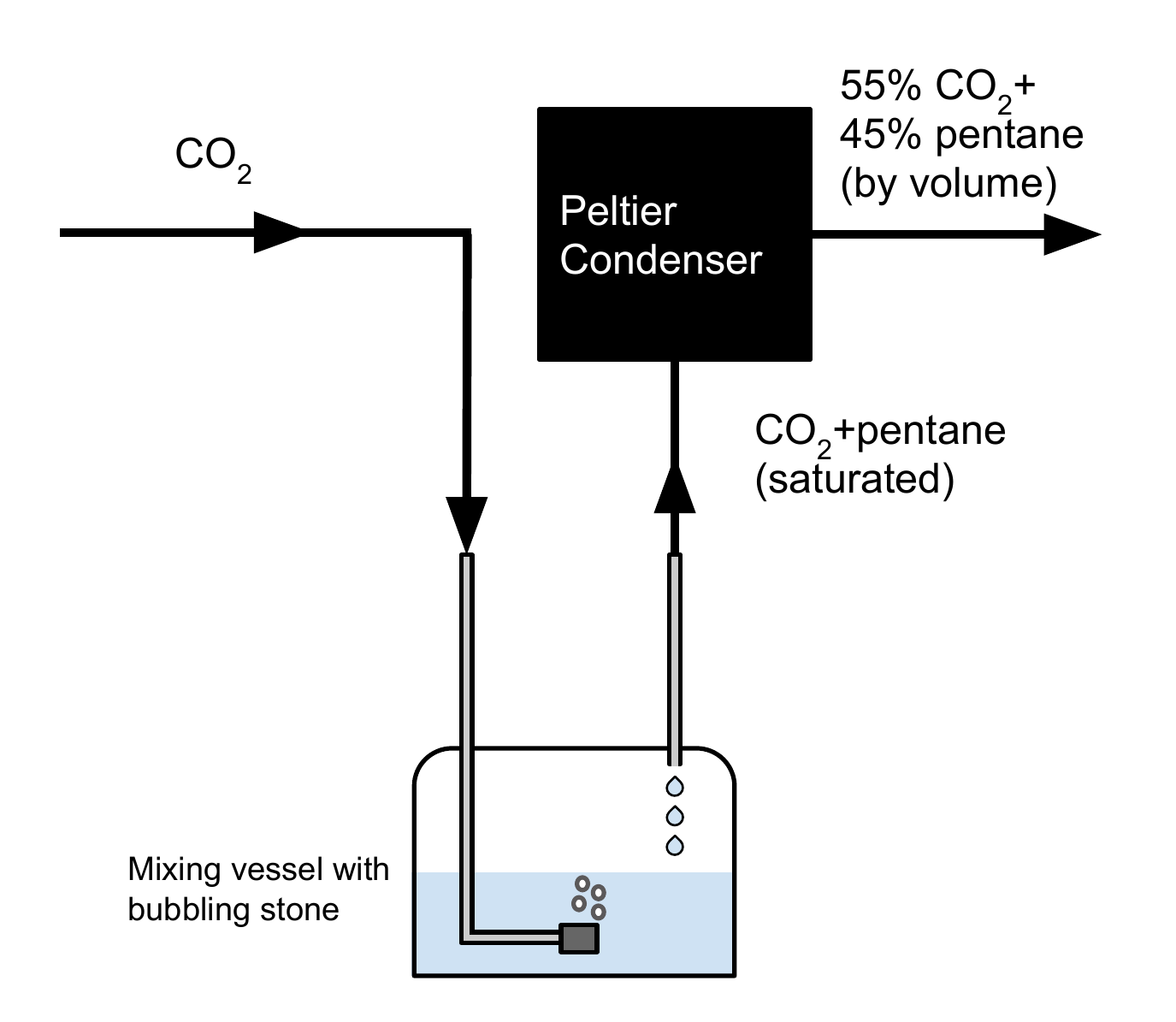}
  \caption{Diagram of the mixing apparatus. The path of the flowing gas is indicated by the arrows. The blue-filled droplets indicate the condensed \np{} which falls back into the vessel by gravity while the round white-filled shapes indicate the bubbling \co{} gas mixing with the liquid \np{}.}
  \label{fig:mixerDiagram}
\end{figure}

The mixing apparatus presented here first creates a saturated \co{}:\np{} mixture, which corresponds to a higher
concentration than desired, at room temperature ($\sim$20\deg{}).
It then cools the resulting gas mixture to achieve
45~\vp{} of \np{}. The apparatus consists of two principal components: the liquid \np{}
mixing vessel and the Peltier thermoelectric cooler (TEC) condenser system,
as seen in Figure~\ref{fig:mixerDiagram}. The Peltier condenser consists of a cooling plate and a condensing pipe assembly described later. 
The condensing pipe assembly is connected to the mixing vessel by a tubing manifold.
The Peltier condenser is positioned above the \np{} reservoir to ensure that the condensed liquid returns to the vessel by gravity. 
This novel design facilitates the refilling and replacement of the \np{} reservoir
during continuous operation of the gas system by cooling the resulting gas mixture instead of the
vessel where the initial mixing occurs~\cite{openshaw}.
The refilling of the liquid \np{} in the reservoir is done using a glass separatory funnel\footnote{Scientific Equipment of Houston, ``separatory funnel with glass stopcock, 500~\ml{}''.} such that no intake of air occurs during the process. 

The \np{} reservoir consists
of a 7.5~L stainless steel pressure vessel\footnote{\mcmaster{} ``portable Stainless Steel ASME-code pressurized liquid dispensing tank'', 9-inch diameter.} fitted with a dip-tube
connection reaching to the bottom of the vessel and a quick-disconnect fitting on the top of the vessel. \co{} flows 
through a set of bubbling stones fixed to the end of the dip-tube immersed in the
liquid \np{}, while the saturated mixture (roughly 57~\vp{} \np{} 43~\vp{}
\co{} at 20\deg{}) flows out
through the connection at the top of the vessel. The gas mixture residing in the
mixing vessel reaches saturation due to the large surface area of the liquid, the use of 
bubbling stones to increase the interaction area of the liquid \np{} with the \co{}, the large \np{} 
vapour volume in equilibrium with the liquid in the reservoir, and the low flow rate which translates 
into a residence time of about 1 hour for \co{} in the vessel for typical conditions with multiple gas lines in use.

\begin{figure}
  \subfigure[]{
     \includegraphics[width=8cm]{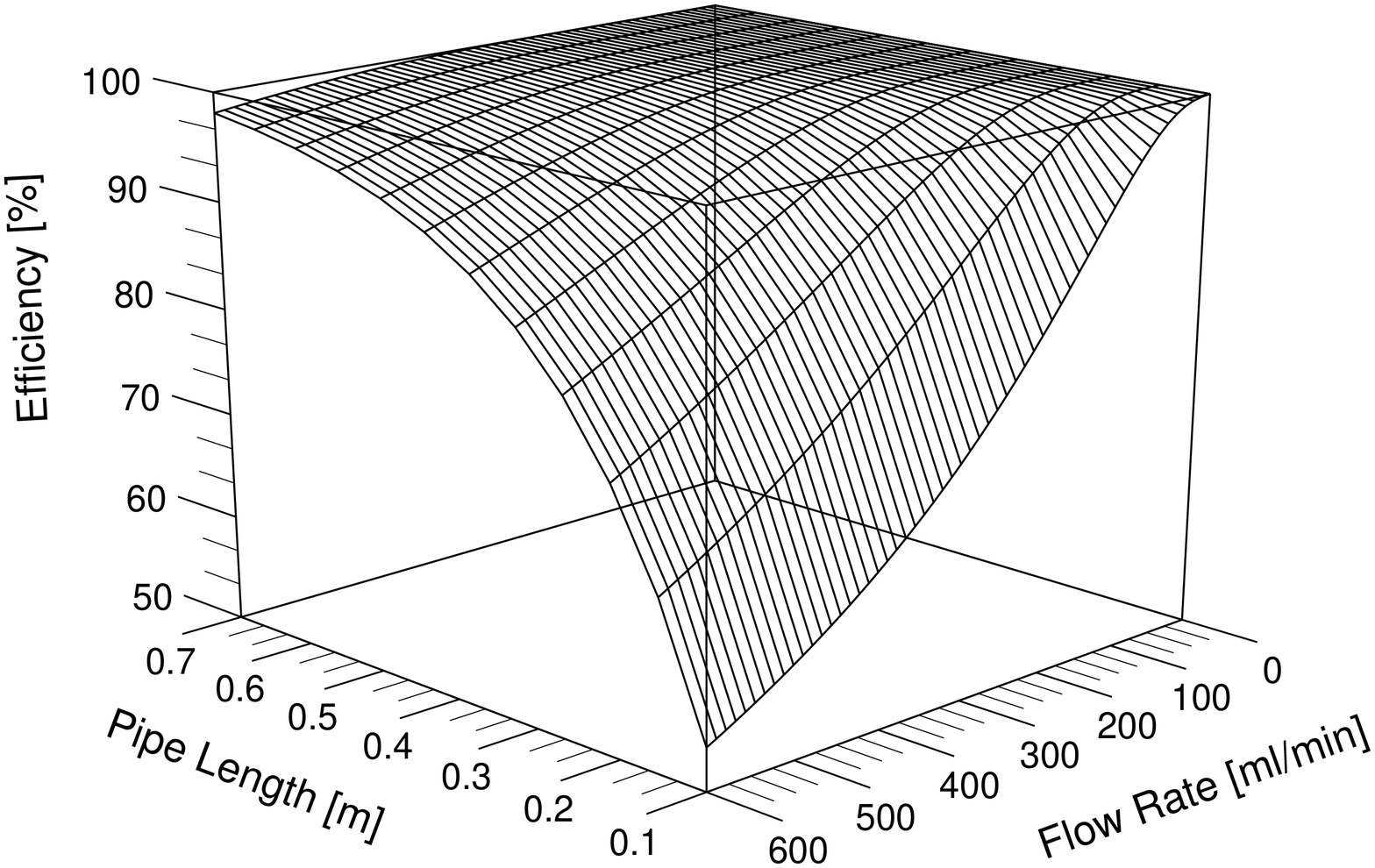}
     \label{fig:GasCalc}
  }
  \subfigure[]{
     \centering
     \includegraphics[width=7.5cm]{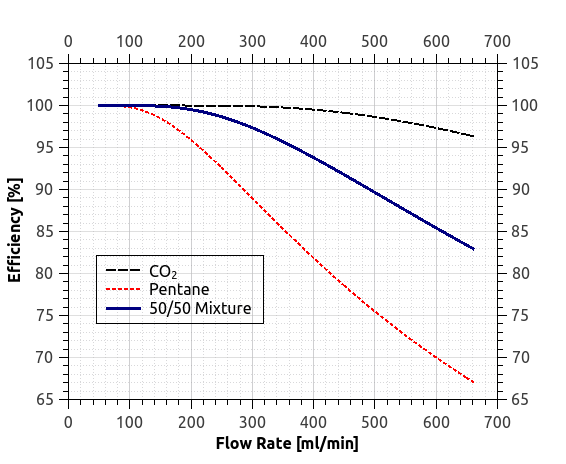}
     \label{fig:GasCalc1D}
  }
  \caption{Cooling efficiency (\( \frac{T_{\textsc{\tiny Out}}-T_{\textsc{\tiny In}}}{T_{\textsc{\tiny Wall}}-T_{\textsc{\tiny In}}}\)) for a single pipe. (a) Cooling efficiency as a function of the volume flow rate and pipe length assuming a 50\%:50\% mixture of \np{} and \co{}. (b) Cooling efficiency assuming a 30~cm pipe length (as is used in the apparatus) for the limiting cases of pure \np{} and pure \co{}, as well as a 50\%:50\% mixture.}
\end{figure}

The Peltier cooling plate, which has a maximum cooling capacity of 200~W at 0\deg{}, was purchased commercially\footnote{TE Technology, Inc. ``CP-200 Peltier-thermoelectric cold plate cooler''.}.
The cooling surface has dimensions 15~cm x 21.5~cm. A simple condensing assembly was
designed to maximise the cooling efficiency by splitting the total
flow to increase the contact area with the Peltier plate and increase the time that the gas spends within the cooling volume. A rudimentary thermodynamic analysis of the cooling of a laminar
gaseous flow in an isothermal pipe indicates that the most sensitive
parameters affecting its performance are the set temperature, the
volume flow rate, and the pipe length~\cite{HeatTransferText}. The calculation is done assuming
fully developed laminar flow in the system, an inner pipe diameter of 1.1~cm, and
literature values for the dynamic viscosity, heat capacity, thermal
conductivity and density of the input \co{}:\np{} mixture. 
The cooling efficiency is defined as ( \(
\frac{T_{\textsc{\tiny Out}}-T_{\textsc{\tiny In}}}{T_{\textsc{\tiny Wall}}-T_{\textsc{\tiny In}}}\)), where $T_{\textsc{\tiny Out}}$ and $T_{\textsc{\tiny In}}$ are the temperatures of the outgoing and incoming gas, while $T_{\textsc{\tiny Wall}}$ is the temperature of the isothermal pipe. 
Figure~\ref{fig:GasCalc} displays the resulting cooling efficiency as a function of the flow
rate and pipe length assuming a 50\%:50\% \co{}:\np{} mixture. 
Figure \ref{fig:GasCalc1D} shows the cooling efficiency assuming a pipe length of 30~cm, as is used in the apparatus, for the limiting cases of pure \co{} and pure \np{}, as well as for a 50\%:50\% mixture that is similar in composition to the actual gas mixture used in the Peltier condenser.
	
\begin{figure}
  \center
  \includegraphics[width=8cm]{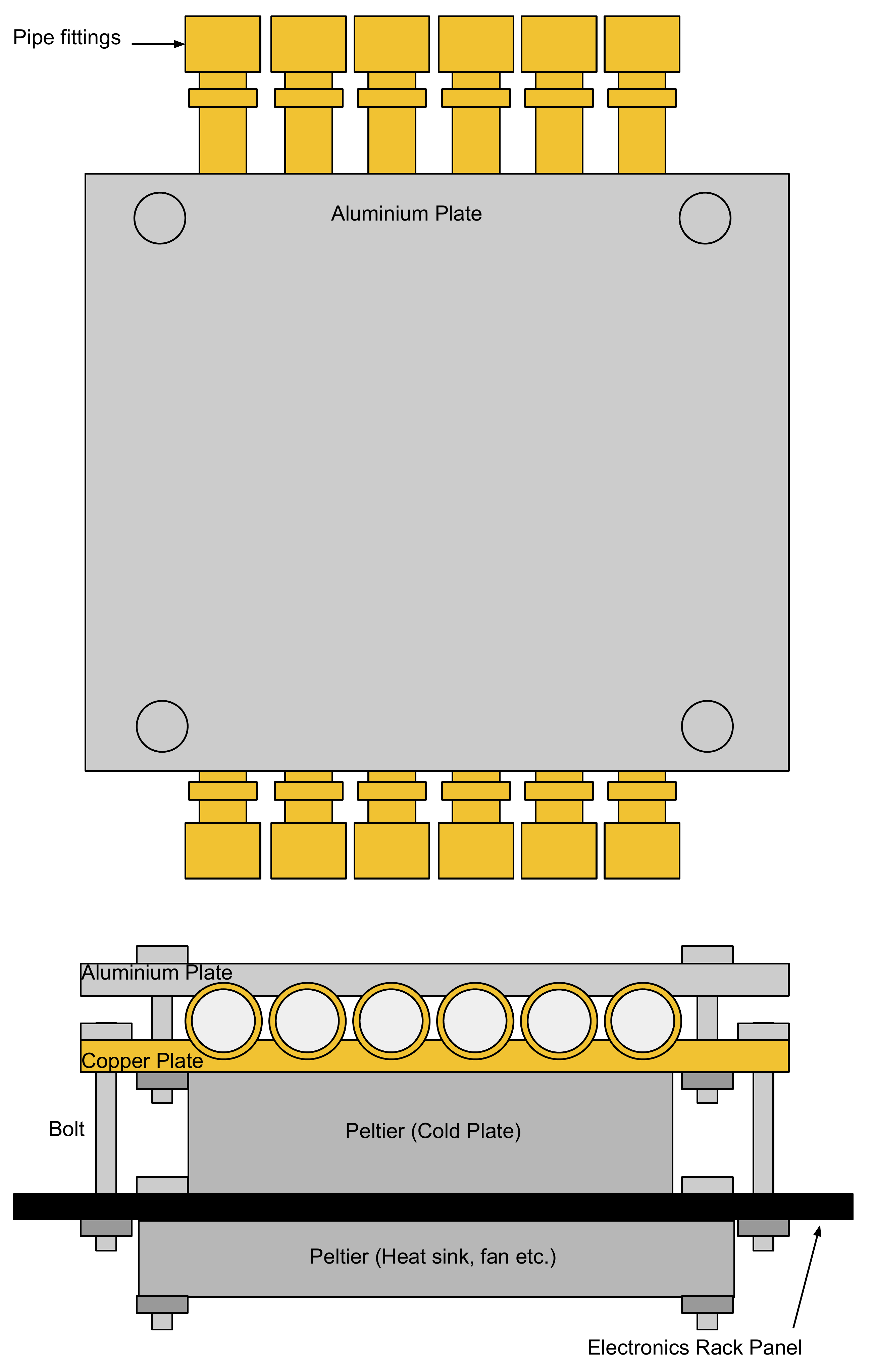}
  \caption{Front and bottom views of the cooling plate and pipe assembly.}
  \label{fig:Condenser}
\end{figure}

The maximum flow rate of gas mixture anticipated for this apparatus is
approximately 525~\ml{}/min, assuming all five \co{}:\np{} mixture distribution lines are in simultaneous use. By dividing the flow in the condensing assembly into
six pipes, the flow in each pipe becomes 87~\ml{}/min. At this flow
rate, as shown in Figure~\ref{fig:GasCalc1D}, employing a pipe length of 30~cm ensures that the mixture should approach the set point of the condenser temperature even when assuming the limiting case of pure \np{} vapour. Thus, considering the simplistic nature of this analysis and implicit assumptions, the choice of using a 30~cm assembly provides a safety factor towards achieving the desired gas temperature and, in turn, the desired \np{} removal by condensation. 
The adopted assembly design consists of
six 1.3~cm outer diameter copper pipes (1.1~cm inner diameter), each 30~cm in length, arranged with a pitch of
3.2~cm to accommodate their fittings. The pipes are then inset and
soldered onto a 25.5~cm x 23~cm x 0.64~cm copper plate. The insets in the
copper plate are machined to a 1.3~cm diameter to match the outer
diameter of the pipes. This assembly is then placed in thermal
contact with the cooling plate of the Peltier condenser and bolted in place. An aluminium plate is added and bolted in front of the pipes to provide cover and mechanical support. Figure~\ref{fig:Condenser} shows a basic schematic for this setup. The main
advantages of this design are that it is simple, requires minimal
machining, is easy to assemble making use of standard and readily available parts, and makes optimal use of the total Peltier cooling plate area.

\begin{figure}
  \center
  \includegraphics[width=15cm]{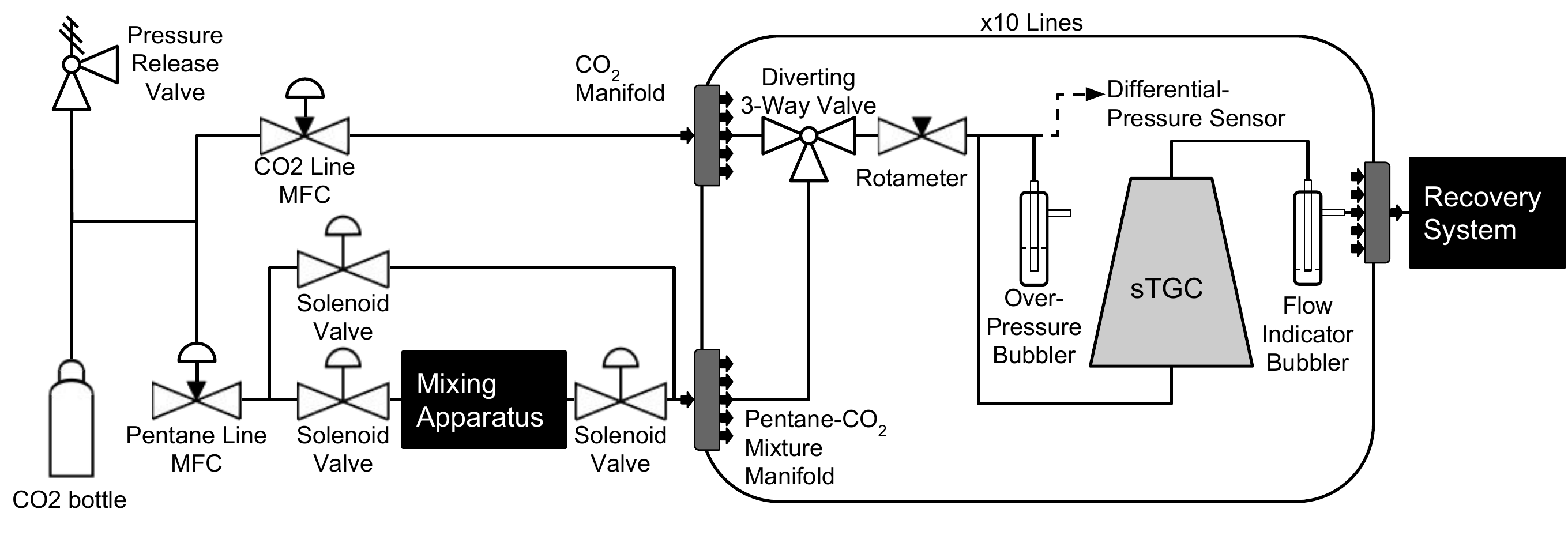}
  \caption{Gas System Diagram. The gas flows from left to right. Two independent Mass Flow Controllers (MFC) control the gas flow input for the two sets of lines (pure \co{} and \co{}:\np{} mixture). The layout for each individual gas line is shown in the box, with the sTGC detector connected to the gas line. In the special case of the dedicated \co{} lines, the 3-way valve is a simpler 2-way valve and the exhaust manifold, situated downstream from the flow-indicator bubblers, is not needed as there is no recovery system.}
  \label{fig:gasDiagram}
\end{figure}

The gas system design is displayed in Figure~\ref{fig:gasDiagram}. The gas flow 
starts with a \co{} cylinder that is set to an output pressure of 70~kPa\footnote{All pressures in the gas system are measured with respect to the atmospheric pressure.}, suitable for operation
of the two mass flow controllers (MFC) directly downstream.
A pressure release valve is placed between the \co{} tank and the MFCs and is set at 103~kPa to
protect the gas system and the \tgc{}s from any accidental sudden increase in pressure. 
One MFC\footnote{MKS ``mass flow controller \co{} M100B'' (M100B02513CS1BV) with power supply 246B.} regulates the flow rate of the pure \co{} input line,
while the other\footnote{Omega ``economical gas mass controllers with integral display'' FMA5514A.} regulates the flow rate of \co{} to the \co{}:\np{} mixer. 
The pure \co{} and the \co{}:\np{} mixture
lines then feed into separate manifolds, which, in turn, feed individual
lines controlled by manual three-way ball valves. It is important to note
that using a three-way valve ensures that there is no possibility for
mixing of the \co{} and \co{}:\np{} mixture flows; each line must be set to
\co{}, to \co{}:\np{} mixture, or be closed. Half of the individual gas lines are dedicated to flow 
only \co{}.

\begin{figure}[t]
  \center
  \includegraphics[width=10cm]{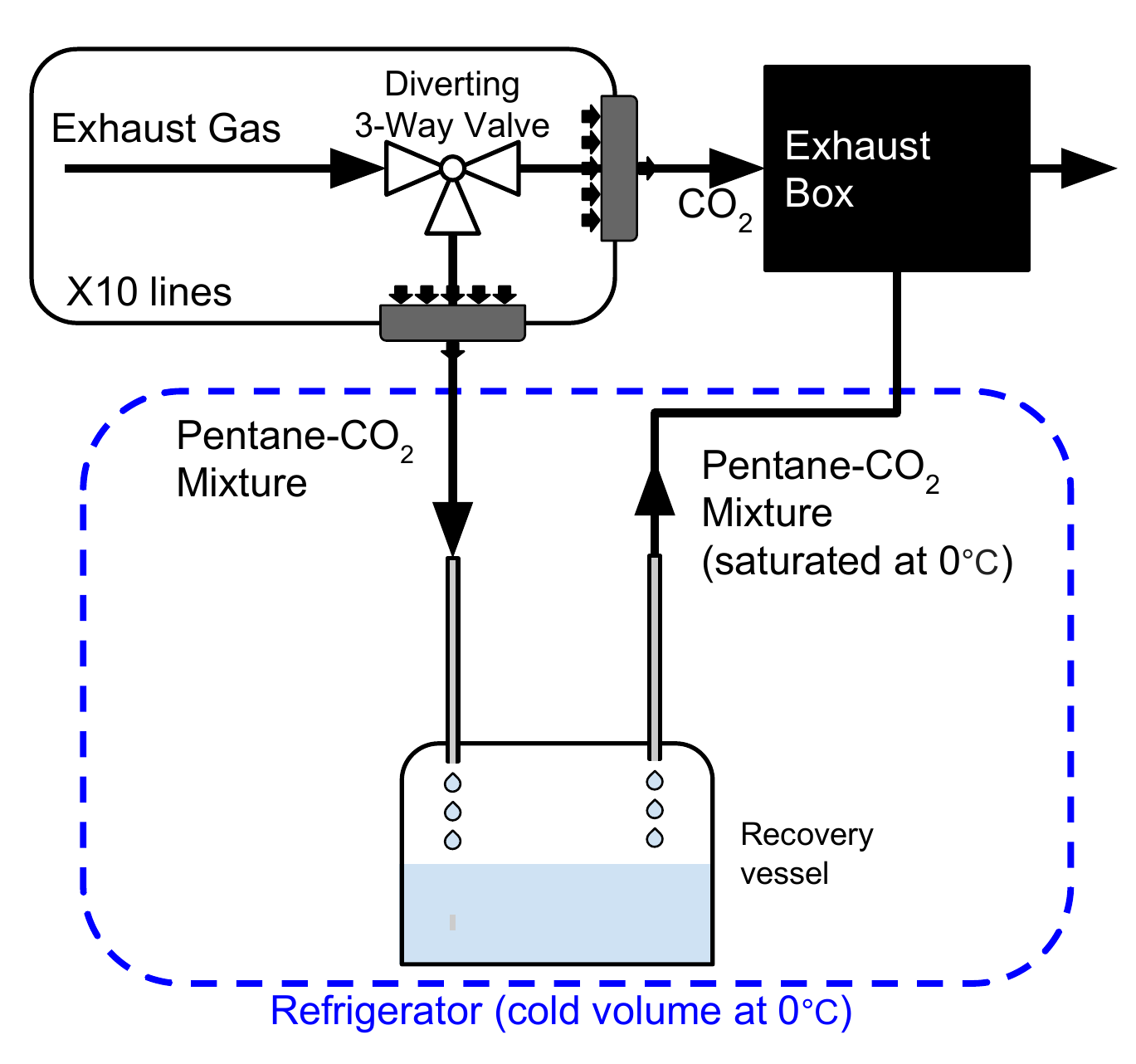}
  \caption{Diagram of the recovery and exhaust apparatus. The path of the flowing gas is indicated by the arrows. The recovery vessel is housed in a refrigerator and maintained at 0\deg{}. The blue-filled droplets indicate the condensed \np{} that falls back into the vessel by gravity. The three-way valve on each line can be used to bypass the recovery system.}
  \label{fig:exhaustDiagram}
\end{figure}

Individual lines consist of a manual needle valve rotameter\footnote{Aalborg model P ``single flow tube meters'' PMR1-015710.}, which is
used to visualise and adjust the flow of the individual line, followed
by two bubblers\footnote{Laboy ``bubbler, mineral oil''.}, one in parallel with the \tgc{} to protect it from mechanical
stresses in the event of an overpressure in the line, and one
in series downstream of the \tgc{} that serves as a visual
indicator that flow is present. A mineral oil bubbler is a device used in a gas circuit to maintain an inert gas environment 
while exercising pressure control. The overpressure bubbler is filled
with a liquid column of vacuum pump oil equivalent to 0.5~kPa in order to 
minimise the differential between the \tgc{} and ambient pressure. The overpressure bubbler is also equipped with a reservoir
on top that allows for all its oil content to be stored in the event of a back-pressure flow, such that no oil could make it
into the gas system tubing and cause a blockage. The
visual indicator bubbler is installed right after the \tgc{}, in series with the individual gas line hence creating a pressure 
drop in the system. It is therefore important to minimise this pressure drop by keeping 
the column height of oil in the bubbler as close to zero as possible, while filling the bubbler enough to show bubbling and to display that there is
flow in the line. 
The flow indicator bubbler is also filled with glass beads immersed in the oil to 
minimise the oil volume exposed to \np{}. Because of its molecular polarity, \np{} accumulates in the
oil and over time raises the overall liquid volume in the bubbler, thereby increasing the
pressure drop of the bubbler. Therefore, minimising the volume exposed to
\np{} by the addition of glass beads in the flow indicator bubbler
minimises the increase in pressure drop of the bubbler over time.
This pressure drop should always remain lower than the overpressure bubbler drop to avoid back flow.
A differential pressure sensor\footnote{Omega ``wet/wet low differential pressure transmitter'' PX154.} is connected in parallel
with the \co{}:\np{} mixture lines for monitoring purposes. This sensor is capable of tracking changes in 
the individual line pressure with respect to the atmosphere as well as transient effects due to, for example, cycling of the building's ventilation system.
Following
the visual indicator bubbler, the flow of the line is then either directed to the
recovery system or directly to the exhaust using a three-way
valve\footnote{\mcmaster{} ``miniature PVC on/off ball valve push-to-connect''.}, as shown in Figure~\ref{fig:exhaustDiagram}. This valve is configured such that its orientation matches that
of the input valve as \co{}:\np{} mixture is typically sent to the
recovery system while \co{} is sent directly to the exhaust. 
The components of the gas system are installed on a 19-inch rack, referred to here as the gas rack.

The recovery system consists of a flammable proof refrigerator\footnote{Fisher Scientific ``explosion proof undercounter refrigerator'' 3556FS.} set to its
coldest setting (0\deg{}), housing a condenser tube connected to a
recovery vessel\footnote{Same type as mixing vessel, same diameter, but smaller capacity.}. 
The recovery vessel collects re-liquefied \np{} prior to the exhaust.
As displayed in Figure~\ref{fig:exhaustDiagram}, as the \co{}:\np{} mixture flows through the tube it
cools and condenses \np{} into the vessel. Approximately 50~\vp{} of the \np{} is recovered at nominal temperature and pressure, as expected from the \np{} fraction curve shown in Figure~\ref{fig:PentaneFraction}.
A large acrylic box, with a volume of approximately 0.34~m$^3$, is used to dilute the remaining \co{}:\np{} mixture with air 
to well below the threshold for flammability prior to the mixture being sent to the exhaust and vented from 
the building. 
The dilution box also decouples the \tgc{} gas system from the negative pressure of the exhaust system.  
Diluting the mixture also reduces the risk of liquid \np{} condensing at or near the outside vent when outdoor temperatures get substantially lower than the mixing temperature. 

The gas system lines and connections are built using nylon tubing\footnote{\mcmaster{} ``extra-flexible nylon tubing .180" ID, 1/4" OD, .035" wall thickness, semi-clear white (also opaque black)'', 50 ft. length (5112K63); ``extra-flexible nylon tubing .275" ID, 3/8" OD, .050" wall thickness, semi-clear white'', 100 ft. length (5112K65).},
nylon push-to-connect\footnote{\mcmaster{} ``push-to-connect tube fittings''.} couplings and acetal
quick-disconnect\footnote{\mcmaster{} ``air and water quick-disconnect tube couplings''.} couplings with automatic shutoff
valves. These components, while exhibiting the necessary material
compatibility, are also easily assembled and interchanged, making the
system simple and convenient to manipulate and modify. Rotameters on
all ten lines have a maximum measurable \co{} flow of 90~\ml{}/min (they can flow up to 100~\ml{}/min, beyond the displayed readings), with the exception of a larger rotameter installed on one
of the \co{} dedicated lines, which has a maximum measurable \co{} flow of 730~\ml{}/min. The larger rotameter allows
flexibility in the number of quadruplet modules that can be flushed with
\co{} at any given time. The mass flow controllers are capable of flowing
0-1000~\ml{}/min of \co{}, ensuring that in the event of a
full system flush, 100~\ml{}/min can be flushed through each line simultaneously. The
MFC for the pure \co{} lines is also equipped with a manual override switch such that an
operator can always bypass the slow-control system (see Section~\ref{slowcontrol}) and flush the system manually with \co{}. 
Solenoid valves\footnote{ASCO 8215G010 24VDC.} 
are installed in specific locations (see Figure~\ref{fig:gasDiagram}) and
are directly actuated by the control system. The valves are closed in the absence of current,
stopping gas flow in the event of a loss of power. Temperatures of the solenoid valves are used as indications of their recent prevailing energized or de-energized states.
Explosive gas sensors provide analog readings and have relays hard-wired to the solenoid valves and the HV/LV crate, 
which is used to trip off the system and isolate the \np{} vessels in the event of a flammable gas leak.

\subsection{Characterisation of the Gas System}

The mixing procedure provides an adjustment from
the concentration of \np{} in \co{} at ambient temperature saturation (57~\vp{} \np{} at 20\deg{}) to the desired concentration
of 45~\vp{} using a Peltier condenser. In order to determine
the condenser temperature set point needed to achieve this value, two
different methods are used to characterise the mixture. The
characterisation methods are also useful to understand the evolution
of the concentration at various points within the distribution system.

One method makes use of the knowledge of the mass of \np{} 
and volume of \co{} consumed in the run to estimate the mixture volume fraction.
The amount of \np{} consumed during the run is determined by weighing the \np{}
reservoir before and after the run.
The volume of \co{} is estimated by integrating the flow rate of the MFC
over the period of interest. The uncertainty on the resulting volume fraction is calculated 
by propagating the error of the mass scale ($\sim10\%$) and the MFC measurements ($\sim0.02\%$).

\begin{figure}
  \center
  \includegraphics[width=10cm]{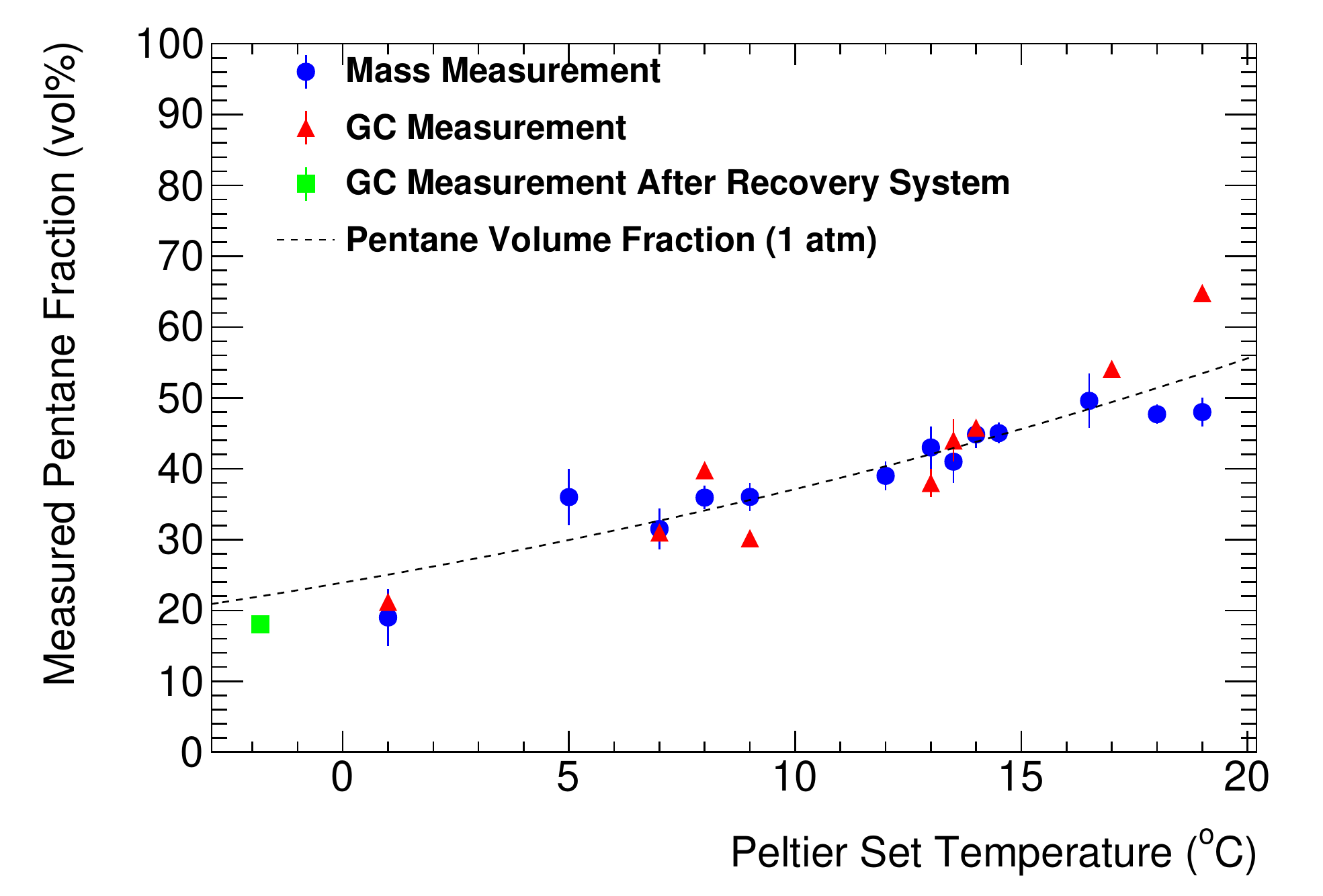}
  \caption{The \np{} concentration of the gas mixture as produced by
    the mixing apparatus, measured using two different methods: a mass
    measurement (blue points) and a gas chromatography (GC)
    measurement (red points). Multiple measurements for each Peltier
		set point are combined. The point in green
    indicates the GC measurement of a gas sample collected after the
    recovery refrigerator. The dashed line shows the theoretical calculation
    for the \np{} vapour pressure, adjusted for the gas temperature measured inside the Peltier condenser.}
  \label{fig:Peltiercurve}
\end{figure}

For a selection of runs where the above method is used, small gas
samples were also collected to be later analysed using gas
chromatography (GC)~\cite{GC} with a Thermal Conductivity Detector
(GC-TCD). Helium is used as the carrier gas in the GC column to obtain
the best sensitivity to \np{}. The GC-TCD allows the determination of the
quantity of each gas type contained in a given sample. 
A gas mixture with a known concentration\footnote{Praxair, Custom specialty gas mixture, CGA-510 connector, CD PT10C-FX.} of 10~\vp{} of
\np{} in \co{} (with a relative tolerance of $\pm$5\%) is used to provide a calibration for the GC-TCD.
Two calibration curves are obtained from the same calibration sample and are used to determine the uncertainty on
the measurement. Gas samples are collected from the gas system at the
point where the \tgc{} would normally be connected. Additional samples
are collected at the output of the \np{} recovery
system to verify its performance as a function of the refrigerator set point. Each sample is then analysed with the GC-TCD.
Uncertainties due to the GC-TCD are evaluated by using different calibration results.  
The concentration uncertainty is computed
as the difference between the extremum values and the central value at
the middle of the interval. Additional systematic uncertainties
associated with laboratory manipulations are not evaluated. The results are shown in
Figure~\ref{fig:Peltiercurve} where the sample concentrations are plotted as a function
of the Peltier condenser set point temperature, and multiple measurements for each Peltier
set point have been combined.
The expected \np{} vapour pressure curve, adjusted for the gas temperature measured inside the Peltier condenser, is shown overlaid on
the data, showing that the measured concentrations as a function of
temperature exhibit the expected trend. 

\begin{figure}
    \centering
    \subfigure[]{
        \centering
        \includegraphics[width=0.5\textwidth]{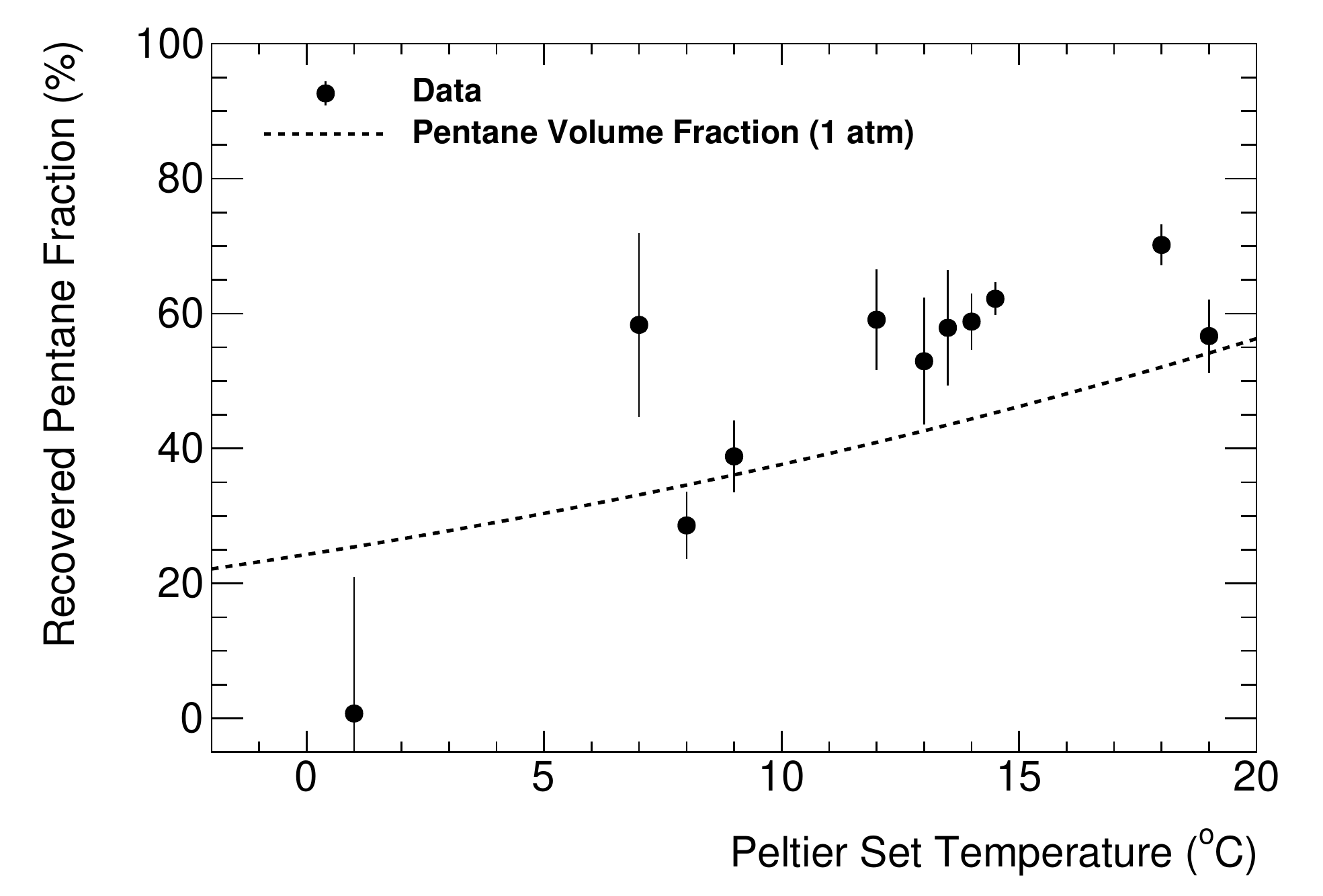}
				\label{fig:RecoveryEfficiency}
    }%
    ~ 
    \subfigure[]{
        \centering
        \includegraphics[width=0.5\textwidth]{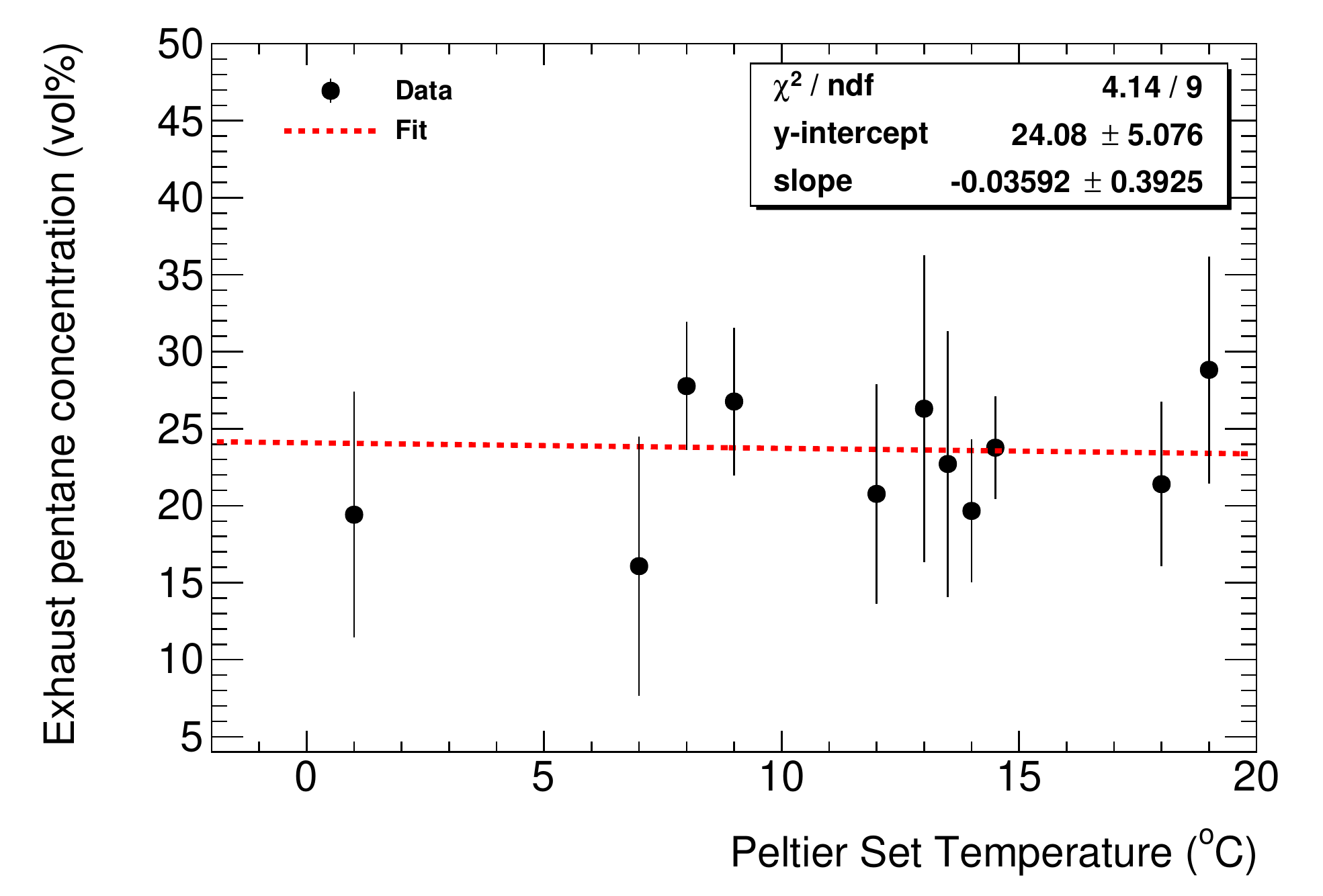}
				\label{fig:ExhaustPentaneConcentration}
    }
		\caption{Characterisation of the \np{} recovery system. (a) The recovered fraction of the \co{}:\np{} gas mixture in the
    \np{} recovery vessel in the refrigerator downstream of the \tgc{} and (b) the exhaust \np{} concentration 
		downstream of the \tgc{} and refrigerator for different data taking runs. In (a), the points roughly align on the pentane volume fraction curve (dotted line) 
		while in (b), the points are fitted with a straight line (red dotted line). Multiple measurements for each Peltier
    set point are combined.}\label{fig:ExhaustCharac}
\end{figure}

The recovered fraction, defined as the fraction of \np{} in the gas mixture collected by the recovery system over the duration of a run, is
measured by taking the ratio of the mass change in the \np{} mixing vessel and its contents
to that in the \np{} recovery vessel and its contents. The uncertainty in this 
measurement is calculated by propagating the uncertainty on the mass scale ($\sim$10\%). It is 
found that the recovered fraction is greater for higher input \np{} volume fractions 
(i.e.~warmer Peltier settings) and colder refrigerator settings, as expected. For typical running conditions, i.e. $\sim$14.5\deg{} Peltier condenser and 0\deg{} recovery refrigerator settings, the 
recovered fraction is measured to be approximately 50\%, as expected by design of the recovery system. Figure~\ref{fig:RecoveryEfficiency}
displays the measured recovered fraction as a function of the Peltier set temperature, 
for a constant refrigerator temperature of 0\deg{}. 
The exhaust \np{} concentration is measured by using the mass changes in the contents of the 
\np{} reservoir and the recovery vessel, and the volume 
of \co{} consumed in the run. It is given by 
$\frac{\left(V_{\textrm{\scriptsize m}}-V_{\textrm{\scriptsize r}}\right)}{V_{\tiny \co{}}+\left(V_{\textrm{\scriptsize m}}-V_{\textrm{\scriptsize r}}\right)}$, 
where $V_{\textrm{\scriptsize m}}, V_{\textrm{\scriptsize r}}$ and $V_{\tiny \co{}}$ 
are the pentane evaporation gas volume out of the mixing vessel, the pentane
condensation gas volume in the recovery vessel and the integrated gas flow though the MFC, respectively, treating the 
pressure and temperature as constant during the whole run. 
Given a constant refrigerator temperature, the behaviour of the recovered pentane fraction is driven by the Peltier set point and should follow the pentane volume fraction curve, which is shown in Figure~\ref{fig:RecoveryEfficiency} as a dashed line. 
Figure~\ref{fig:ExhaustPentaneConcentration} 
shows the average exhaust concentration of \np{} downstream from the refrigerator
for each run, as a function of the Peltier temperature. No significant dependence is 
observed, implying the exhaust \np{} concentration is determined entirely by the refrigerator
temperature rather than the Peltier set temperature, as expected. 
Multiple measurements for each Peltier set point are combined in both plots of Figure~\ref{fig:ExhaustCharac}.
The exhaust mixture (with a \np{} concentration of about 20\%) is later diluted in air to a concentration that is well below the lower explosive limit (LEL) for \np{}, as verified by an explosive gas sensor placed just prior to the exhaust for the purpose of this measurement.

\section{Slow-Control System}\label{slowcontrol}

The operation of the \tgc{} detectors in the testing facility must 
ensure both the integrity and safety of the \tgc{}s and
that all measures are taken to minimise possible sources of human error.
Such a facility can exhibit different types of hazards: 
risk of injury to personnel, damage to the laboratory, damage to the \tgc{} 
detectors due to problems with the gas system, and damage to the \tgc{} detectors due to environmental conditions.
A slow-control system and its associated state
machine~\cite{Bitter} have been developed at McGill to safely operate the facility by providing automated safety actions, system control and conditions monitoring. 
The system is tailored to the requirements of a testing facility in a university laboratory and is therefore distinct from the global slow-control paradigm used for the ATLAS experiment at CERN.

The slow-control system incorporates environmental sensors to monitor the ambient laboratory conditions as well as sensors that monitor the individual gas system components' conditions and performance.
The output values of these sensors can trigger safety actions, such as bypassing
the \np{} mixer and recovery refrigerator, or turning off the high/low voltage supply, 
through either hard-wired relay circuits or software controls. The
data from each sensor are logged, organised and presented to the operator
through graphical interfaces.
Since the testing facility operates with or without
the presence of an operator, the state machine includes an alert
system to inform personnel, via email and SMS messages, when a warning or error condition is raised, as well as a
remote monitoring system publishing the current state of the system in real-time on a web page.

%

The state machine 
is the primary user
interface, with built-in protocols for the gas system and high/low voltage system
operation.  It
imposes conditions for the flushing with \co{} prior to \co{}:\np{} mixture flow
and ensures that HV is only applied once a stable gas composition has been reached inside
the detector volume. The slow-control system includes
hard-wired safety actions that take place when an error condition is raised by the
sensors, independently of the state of the slow-control system software or the state machine.
It is also designed to halt operation and revert to a safe state in the event of a power loss or loss of connectivity.

\subsection{Design and Implementation}

The \tgc{} state machine is a software interface programmed using \labview{}. It controls,
monitors and operates the gas flow, the high voltage supplied to the
\tgc{}s and the low voltage supplied to the readout electronics. The state machine sequentially guides the operator through the
procedure needed to operate these systems and to perform a successful
run, while concurrently displaying the current state of the system and
imposing fail-safe operation in the event of a failure.

\begin{figure}
        \includegraphics[height=8cm]{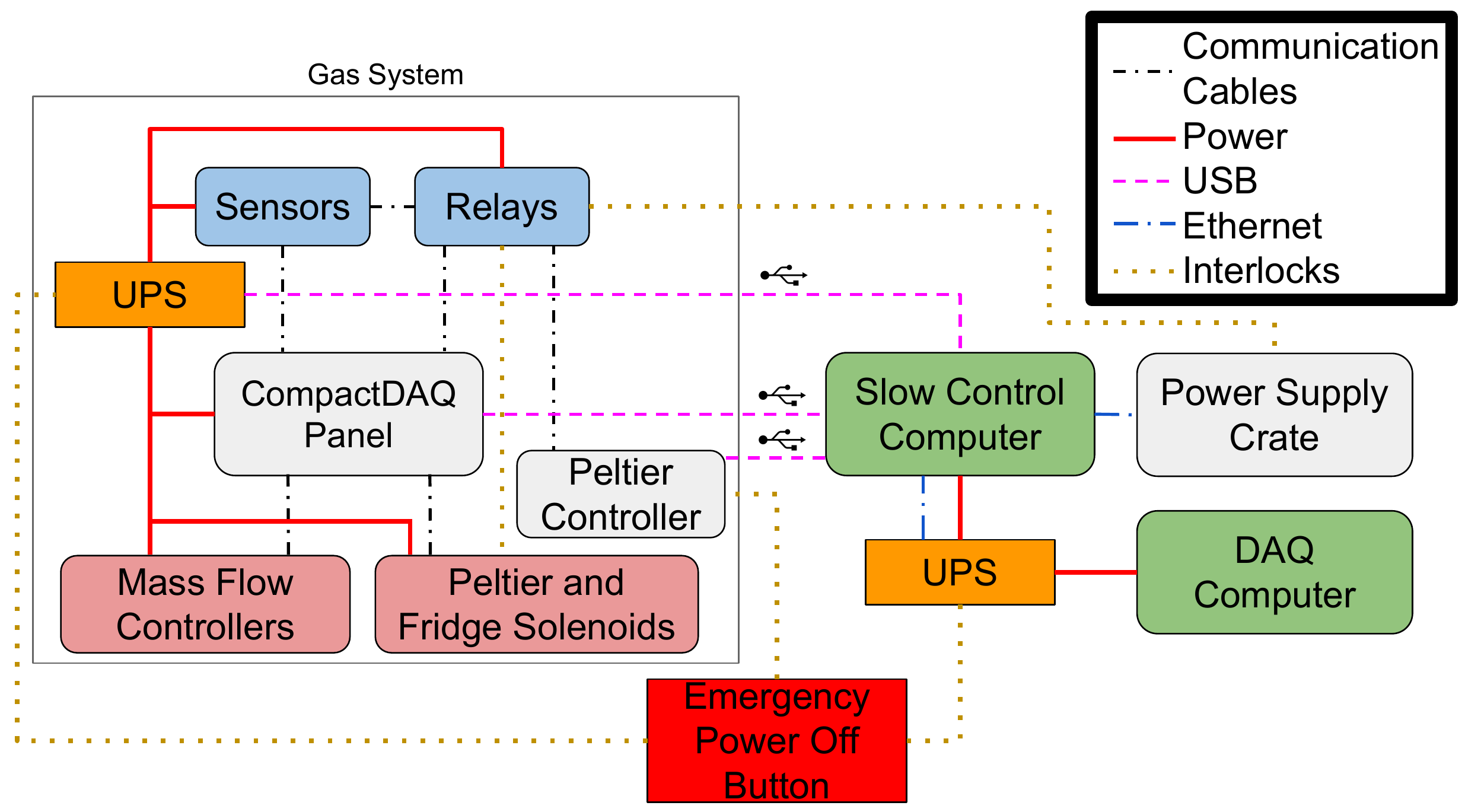}
	\caption{Diagram of the connection between the different slow-control components. The interlocks from the ``Relays'' box are those activated by the emergency relay.}
	\label{fig:ConnectionDiagram}
\end{figure}

The sensor monitoring and control signal output is implemented using a National Instruments CompactDAQ system, which is composed of a crate (readable via a single USB cable) and six modules, listed in Table~\ref{tab:NImodules}. Four of the modules read sensor inputs and two provide signal outputs. The specific sensors monitored by the CompactDAQ system, excluding temperature and ambient condition sensors, are listed in Table~\ref{tab:sensors}. Output voltage signals provide the set points for the two MFCs, controlling the gas flow in the system. The relay output controls a separate emergency relay. The emergency relay is triggered either by the software when the slow-control system detects problematic operating conditions or directly via the hardware, bypassing the slow control system, if either the explosive gas sensors or the exhaust flow sensor detect dangerous operating conditions. The emergency relay itself cuts power to the solenoid valves, thus isolating and bypassing the pentane mixer and pentane recovery system. It also triggers the interlock on the high-voltage source, the CAEN power supply crate that hosts positive and negative polarity HV and LV cards. The interlock signal is sent via the front panel Lemo connector on the CAEN power supply, which ramps down the HV at
the fastest possible rate (measured to be 200 V/s). Control and monitoring elements external to the CompactDAQ system, which include the Peltier temperature controller, the Uninterruptible Power Supply (UPS), and the CAEN HV power supply, are managed using proprietary software that interfaces directly with the LabVIEW environment. The Peltier controller and the UPS are connected via USB while the CAEN crate is connected with ethernet. The UPS is used to power the essential sensor, relay and solenoid-valve elements of the slow-control system in the event of a power cut, long enough (about two hours) that the entire system can be flushed through the \emph{\co{} Bypass} state; see~\S\ref{sec:operations} for a description of this state. An emergency stop button can be used to activate the interlocks on the UPS units, cutting the power to the computers as well as the gas rack, which, in turn, causes the emergency relay to trip. This removes spark-inducing sources and electrical hazards present in the laboratory and in addition to sealing off the pentane vessels, for example, in case of a fire in the laboratory. The general layout of the slow-control connections is shown in Figure~\ref{fig:ConnectionDiagram}.
The Peltier system controller and the HV/LV crate are read into LabVIEW directly via USB and ethernet, respectively, using software provided by the manufacturer.

\begin{table}
  \caption{National Instruments CompactDAQ modules.}
  \label{tab:NImodules}
  \begin{tabular}{|m{1.4cm}|m{4.25cm}|m{4.75cm}|m{4cm}|}
    \hline
    \textbf{Model Number} & \textbf{Description} & \textbf{Input / Output} & \textbf{Associated Components}\\
    \hline
    NI-9203 & Current Input Module & \rr $\pm20$~mA, 200~kS/s, 8-Channel & Gas line pressure sensors;
    Combustible gas detectors \\
    \hline
    NI-9205 & Voltage Input Module & \rr $\pm10$~V, 250~kS/s, 16-bit, 32-Channel & \rr MFCs; \co{} tank pressure; Exhaust sensor; Emergency relay trip; Humidity / Ambient pressure \tn
    \hline
    NI-9213 & \rr Thermocouple Temperature Input Module & \rr $\pm 78$~mV, 75~S/s aggregate, 16-Channel & \rr Solenoid valve temperatures; Gas mixer internal temperature \tn
    \hline
    NI-9217 & \rr PT100 RTD Temperature Input Module & 0~$\Omega$ to 400~$\Omega$, 400~S/s aggregate, 4-Channel & \rr Pentane recovery fridge; Ambient temperature \tn
    \hline
    NI-9263 & Voltage Output Module & \rr $\pm$10~V, 100~kS/s/ch simultaneous, 4-Channel & \rr MFC control \tn
    \hline
    NI-9481 & Relay Output Module & \rr SPST Relay, 60 VDC (1~A) / 250~Vrms (2~A), 4-Channel & \rr Emergency relay; Peltier system fan; CAEN HV power supply interlock \tn
       \hline
  \end{tabular}
\end{table}

\begin{table}
  \caption{Sensors monitored by the CompactDAQ system.}
  \label{tab:sensors}
  \begin{tabular}{|m{2.75cm}|m{3.25cm}|l|m{2.1cm}|m{2.7cm}|}
    \hline
    \textbf{Model Number} & \textbf{Description} & \textbf{Dynamic Range} & \textbf{Output} & \textbf{Purpose}\\
    \hline
    \rr Omega PX154-003DI & \rr Differential Pressure Sensor & $0-750$~Pa & $4-20$~mA & \rr Gas line pressure monitoring \tn
    \hline
    \rr Omega FST1001R & \rr Air Flow Probe Relay & $0-5000$~FPM & \rr 2-Channel, 12~V SPST NO relays & \rr Exhaust flow monitoring \tn
    \hline
    \rr iTrans 7814635-2C212C2 & \rr Combustible Gas Sniffer & $0-100$\%~LEL & \rr $4-20$~mA; 3-Channel, 30~V SPST NO relays & \rr Pentane leak monitoring \tn
    \hline
    \rr McMaster-Carr 3196K2 & \rr Pressure Transducer & $0-1000$~psi & $0-10$~V & \rr  \co{} bottle pressure monitoring \tn
    \hline
    \rr Omega FMA5514A and MKS: M100B & \rr Mass Flow Controller & $50-1000$~sccm & $0-5$~V & \rr Monitoring gas flow \tn
    \hline
  \end{tabular}
\end{table}

The system software is separated into three main LabVIEW Virtual Instruments (VIs):
a data acquisition (DAQ) panel, a HV/LV system control panel,
and an overall state-machine panel. The DAQ panel is configured to read and calibrate
the raw data coming from the sensors at a rate of 1~Hz, making this
information available to all other VIs. 
The system allows live monitoring by concurrently displaying and logging the sensor data on the Slow Control computer, which is periodically backed up to a separate server.
The HV/LV panel allows the operator to control
and monitor the HV/LV system. The current state of the system is
globally made available for the use of other VIs. The state-machine
panel controls the gas system and guides the operator through the
global system operation. The operational flow of the system is
displayed in Figure~\ref{fig:StateTransitionDiagram}, and each state is
described in Table~\ref{tab:StateMachineStatesDescriptions}. Each of the ten individual gas
lines has its own dedicated state machine, with some states
affecting all gas lines. The global state of the system is always the
state of the line that is highest in the operations hierarchy, which increases from the
\emph{Dormant} state up through to the \emph{Run} state.
The state of any individual gas line must be consistent with that of the global state machine;
e.g., changing the global state machine to the \emph{\co{} Bypass} state triggers a change of state
of all the active gas lines.

\begin{table}
\begin{center}
  \caption{Description of possible states for the state machine. The states from \emph{Dormant} to \emph{Run} are individually 
controlled for each gas line. The last two states affect all gas lines.}
  \label{tab:StateMachineStatesDescriptions}
  \begin{tabular}{|l|p{9.5cm}|}
    \hline
    \textbf{State} & \textbf{Description} \\
    \hline
    \emph{Dormant} & Inactive state, no gas flow and high voltage off. \\
    \hline
    \emph{\co{} Flush} & \co{} gas flow at high rate in gas line. High voltage off. \\
    \hline
   \emph{Gas Operation} & Gas changed to \co{}:\np{} mixture in the case of a \np{} line. Gas flow at normal rate. High voltage off. \\
    \hline
   \emph{HV Operation} & Gas type same as \emph{Gas Operation}. Gas flow at normal rate, high voltage on. \\
    \hline
   \emph{Run} & Data acquisition state. Gas type same as \emph{Gas Operation}. Gas flow at normal rate, high voltage on, data acquisition active. \\
    \hhline{|=|=|}
   \emph{\co{} Bypass} & Error state. \Np{} systems isolated, high voltage off, \co{} flow at high rate for active lines (not in \emph{Dormant} state). \\
    \hline
   \emph{Pause} & Operator intervention state. Stops gas flow on all active lines temporarily while keeping high voltage on if present. \\ 
    \hline
  \end{tabular}
\end{center}
\end{table}

\subsection{Operations}\label{sec:operations}
Cosmic-ray data-taking requires that the system reaches the \emph{Run} state, after navigating 
from the \emph{Dormant} state through the other intermediate states. The \emph{\co{} Flush} state ensures 
flushing at a high flow rate until a pre-determined amount of pure \co{} has been flowed 
(depending on the \tgc{} size and gas line length) through the detector. This step is necessary to 
prepare a contaminant-free and oxygen-free environment inside the detector. In the \emph{Gas Operation} state, the flushed \tgc{} 
is flowed with the \co{}:\np{} mixture, at a lower flow rate than in the previous state. 
The operator needs to wait until equilibrium is reached inside the detector, quantified in the software as approximately 10 detector-volume changes.
The \emph{HV Operation} state permits HV to be applied to the trigger system PMTs
and the \tgc{} modules,
necessary for data acquisition, ensuring that only channels (or \tgc{}s) 
associated with a gas line in the state machine, and therefore containing a stable gas mixture, may be powered.

\begin{figure}[t]
    \centering
    \subfigure[]{
        \centering
        \includegraphics[height=8cm]{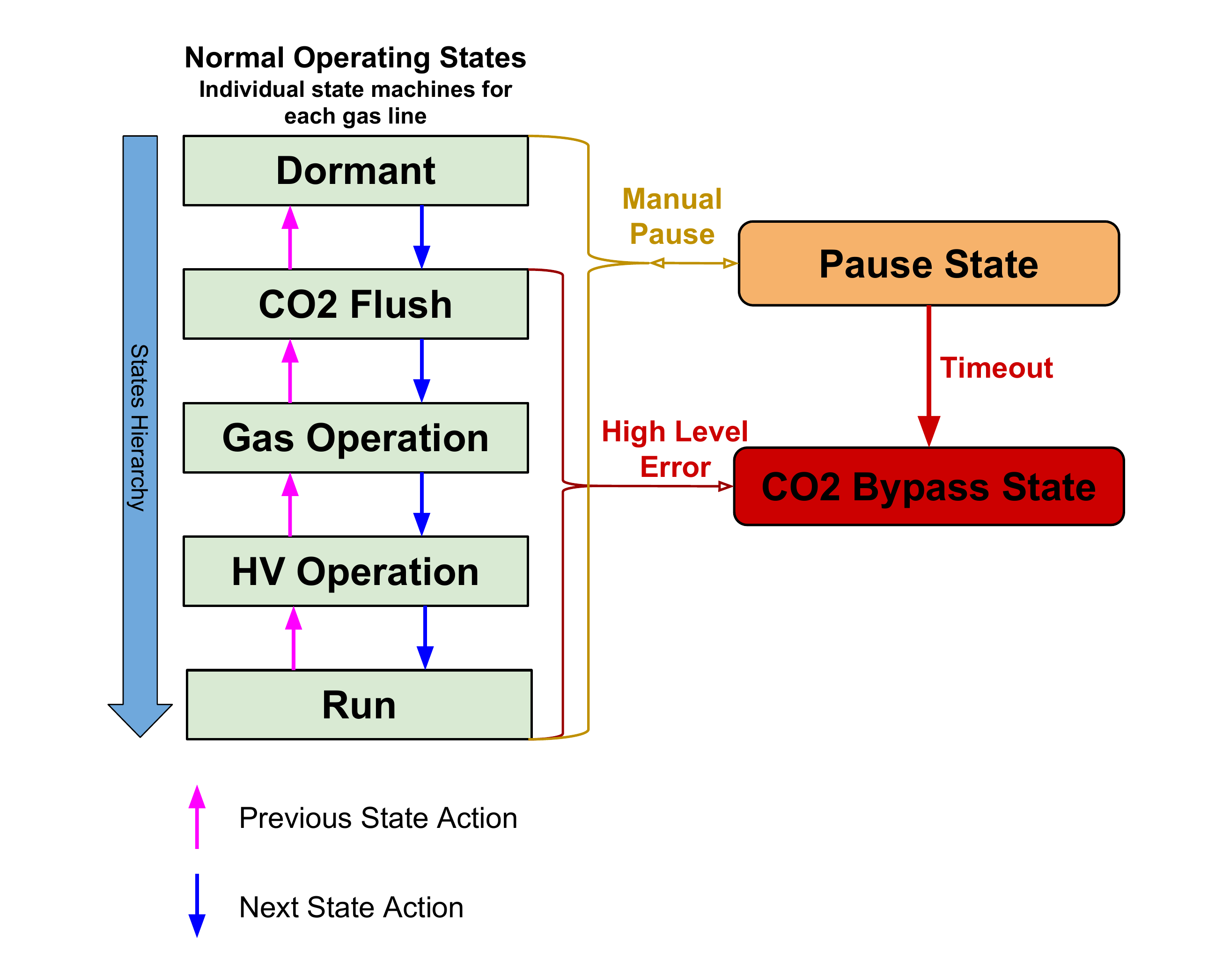}
				\label{fig:StateTransitionDiagram}
    }%
    ~ 
    \subfigure[]{
        \centering
        \includegraphics[height=8cm]{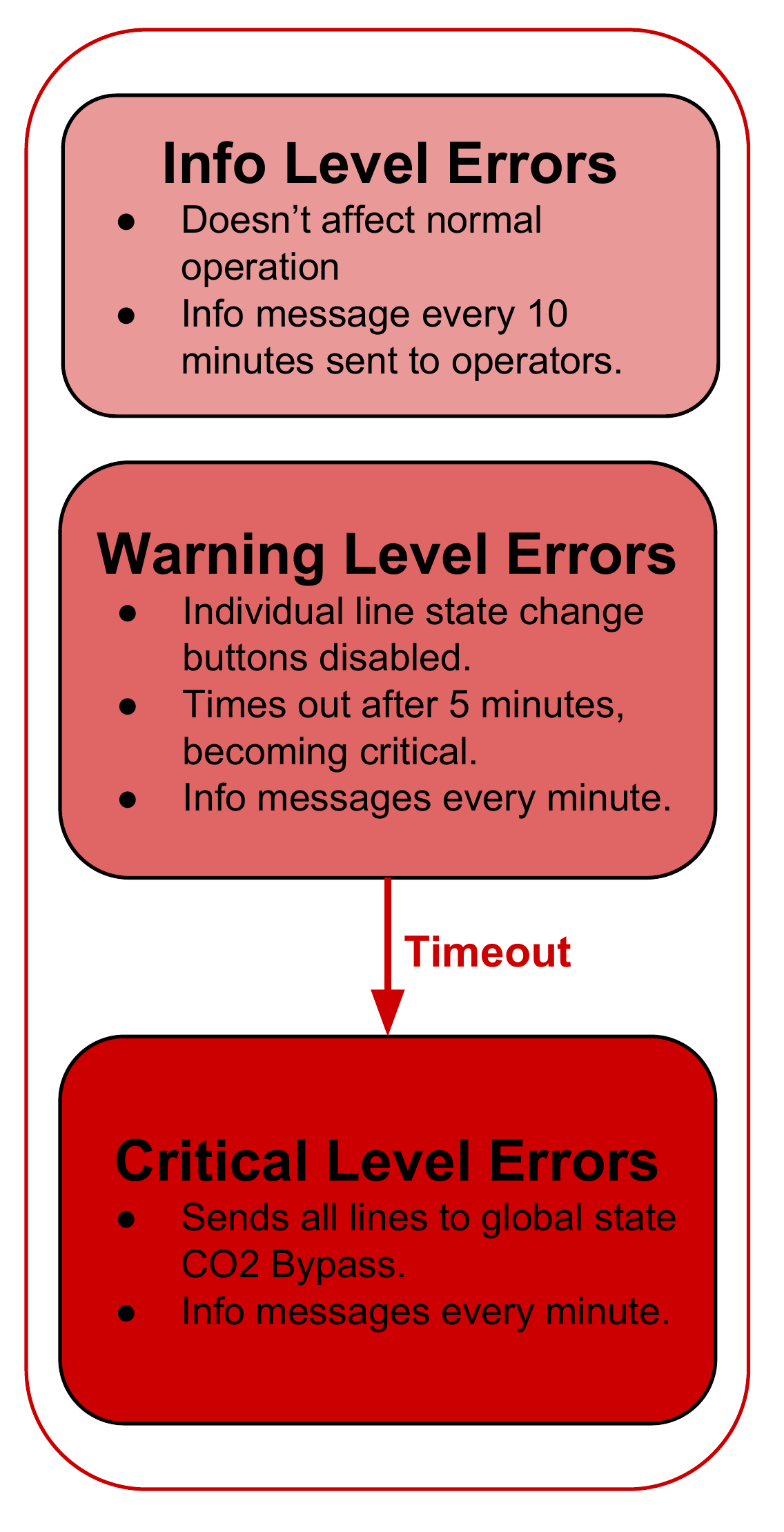}
				\label{fig:StateMachineErrors}
    }
		\caption{State-machine description. (a) State-machine transition diagram. The wide vertical arrow indicates the increasing state hierarchy. The \emph{Timeout} illustrated here is the time allowed for the system to be in the \emph{Pause} state, after which the system transitions to the \emph{\co{} Bypass} state. (b) State-machine errors. The \emph{Timeout} illustrated here is the time allowed for a \emph{Warning level} error to be active before it becomes a \emph{Critical level} error.}
\end{figure}

\begin{table}
  \caption{Possible State Machine Errors}
  \label{tab:StateMachineErrors}
  \begin{tabular}{|p{4.25cm}|l|p{9cm}|}
    \hline
    \textbf{Error} & \textbf{Level} & \textbf{Cause} \\
    \hline
    High \np{} concentration & Critical & A combustible gas sensor detecting a high \np{} concentration around the gas system or in the exhaust chamber. \\
    \hline
    Solenoid valve overheat & Critical & Temperature sensor on one of the solenoid valves detecting an overheat. \\
    \hline
    No exhaust flow & Critical & Flow sensor measuring a critically low flow in the exhaust. \\
    \hline
    Non-optimal exhaust flow  & Warning & Flow sensor measuring a non-optimal flow in the exhaust. \\
    \hline
    Gas flow input mismatch & Warning & Operator input to mass flow controller different than the measured gas flow. \\
    \hline
    High pressure in gas line & Warning & Differential pressure on gas line measuring pressure high enough to potentially damage an \tgc{} chamber. \\
    \hline
    Room temperature low & Warning & Room temperature measured to be low enough to potentially cause \np{} condensation in gas system. \\
    \hline
    Very high room humidity & Warning & Humidity sensor measuring extreme humidity levels. \\
    \hline
    Peltier TEC set temperature mismatch & Info & The temperature of the Peltier TEC controller set by the operator doesn't match its control temperature. \\
    \hline
    Refrigerator not cold & Info & Refrigerator temperature sensor measuring temperatures far above normal operating temperature. \\
    \hline
  \end{tabular}
\end{table}

When actions are required from the operator, the system prompts the operator and waits for confirmation that the action has been taken. 
For example, when moving a given gas line from the \emph{\co{} Flush} to the \emph{Gas Operation} state, the operator is asked to 
turn the input and exhaust manifold valves for the given line from the \co{} to the \np{} position, hence stopping the \co{}
flow and allowing the mixture to circulate. The various
sensors are continuously monitored and alerts are displayed in a dedicated message log window. The errors are 
classified into three levels as shown in Figure~\ref{fig:StateMachineErrors}. \emph{Info level} errors, which correspond to system information messages, 
don't affect normal operation. \emph{Warning level} errors prompt the operator to take some action
before the state can be changed.
If not addressed by the operator within five minutes, \emph{Warning level} errors are promoted to \emph{Critical level}. 
\emph{Critical level} errors cause the system to automatically transition into a predefined safe state, \emph{\co{} Bypass}, which stops all data taking and \np{}
flow while forcing \co{} flow through all lines by setting the appropriate solenoid valves to their open or closed position. 
A list of state-machine errors is shown in
Table~\ref{tab:StateMachineErrors}. If a brief intervention is needed, an operator can manually initiate a pause using the control panel.
The \emph{Pause} state halts all gas flow temporarily on all lines, giving the operator a prompt to resume within 10 minutes. 
Failure to resume operation after the prescribed time has elapsed results in all active lines going to the \emph{\co{} Bypass} state. 
The \emph{Pause} state is used to do small interventions on the gas system, such as swapping pentane reservoirs
or connecting a new \tgc{} to the gas system, without perturbing the gas flow on the already connected lines.

If a \emph{Critical level} or \emph{Warning level} error has been active for more than one minute (10 minutes for \emph{Info level}) and no action was taken by the operator, 
the state machine starts sending alert messages (via email and SMS) to operators periodically until 
an action is taken by the operator to resolve the error.
Time intervals between messages sent of any level are doubled at every iteration until the problem is resolved,
in order to control the inflow of messages to the operators.

\subsection{Performance}\label{sec:performance}
The slow-control system has been used extensively over more than a
year for cosmic-ray data taking with an \tgc{} prototype and has
maintained stable conditions for operations. The performance of the
system is assessed and described in the following section for expected
operations scenarios.

\begin{figure}
    \centering
    \subfigure[]{
        \centering
        \includegraphics[width=0.5\textwidth]{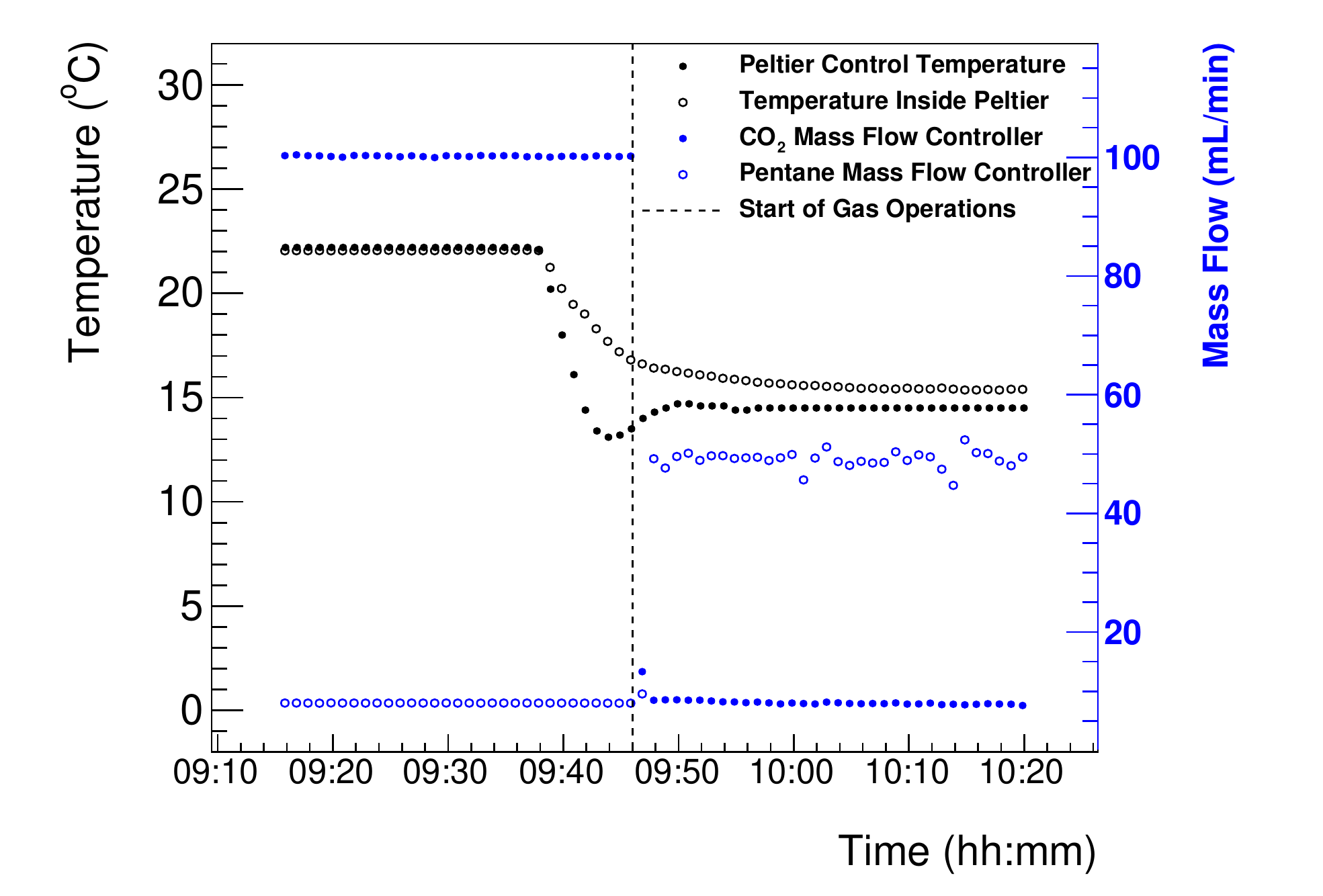}
                                \label{fig:StandardStart1}
    }%
    ~
    \subfigure[]{
        \centering
        \includegraphics[width=0.5\textwidth]{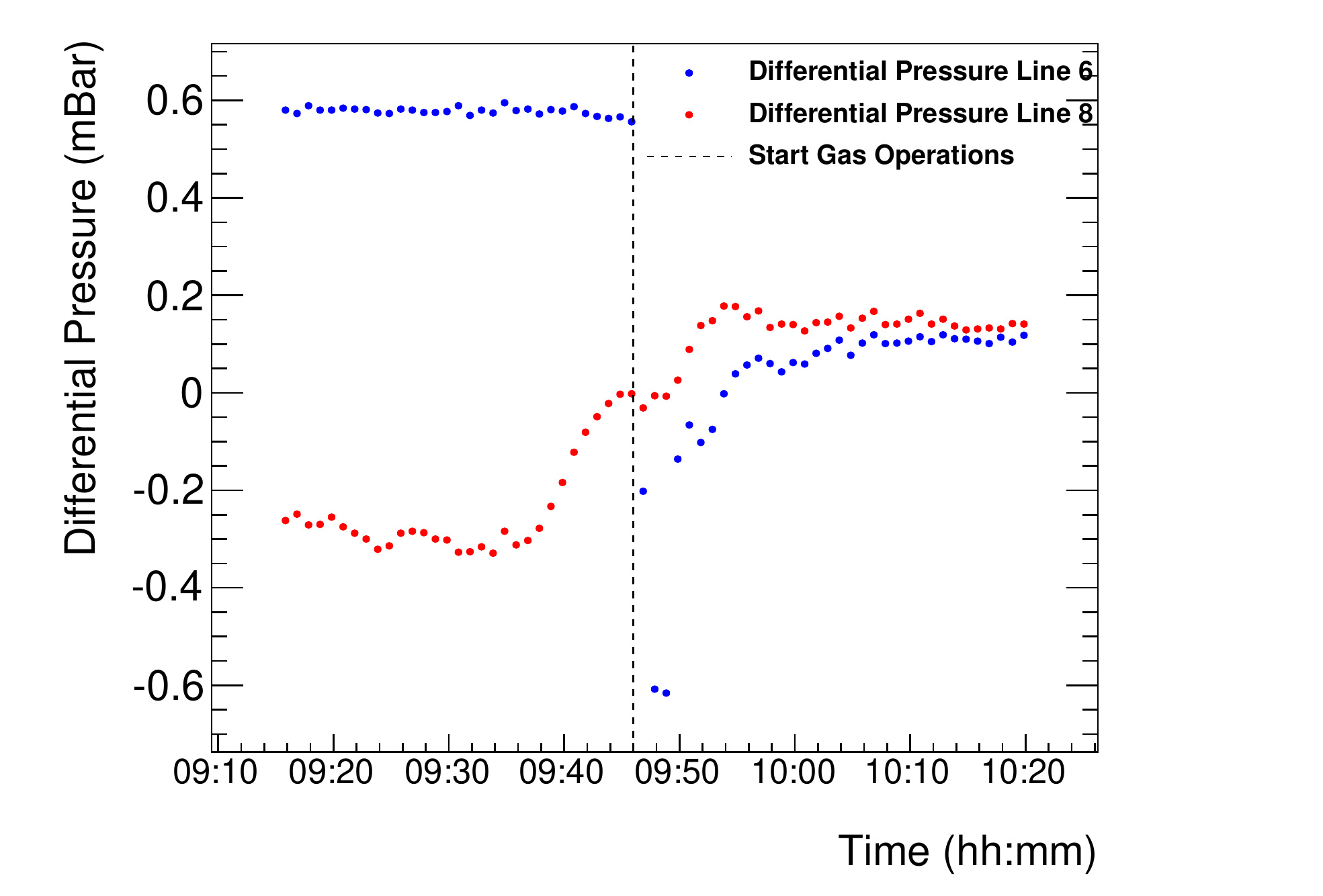}
                                \label{fig:StandardStart2}
    }%
    
    \subfigure[]{
        \centering
        \includegraphics[width=0.5\textwidth]{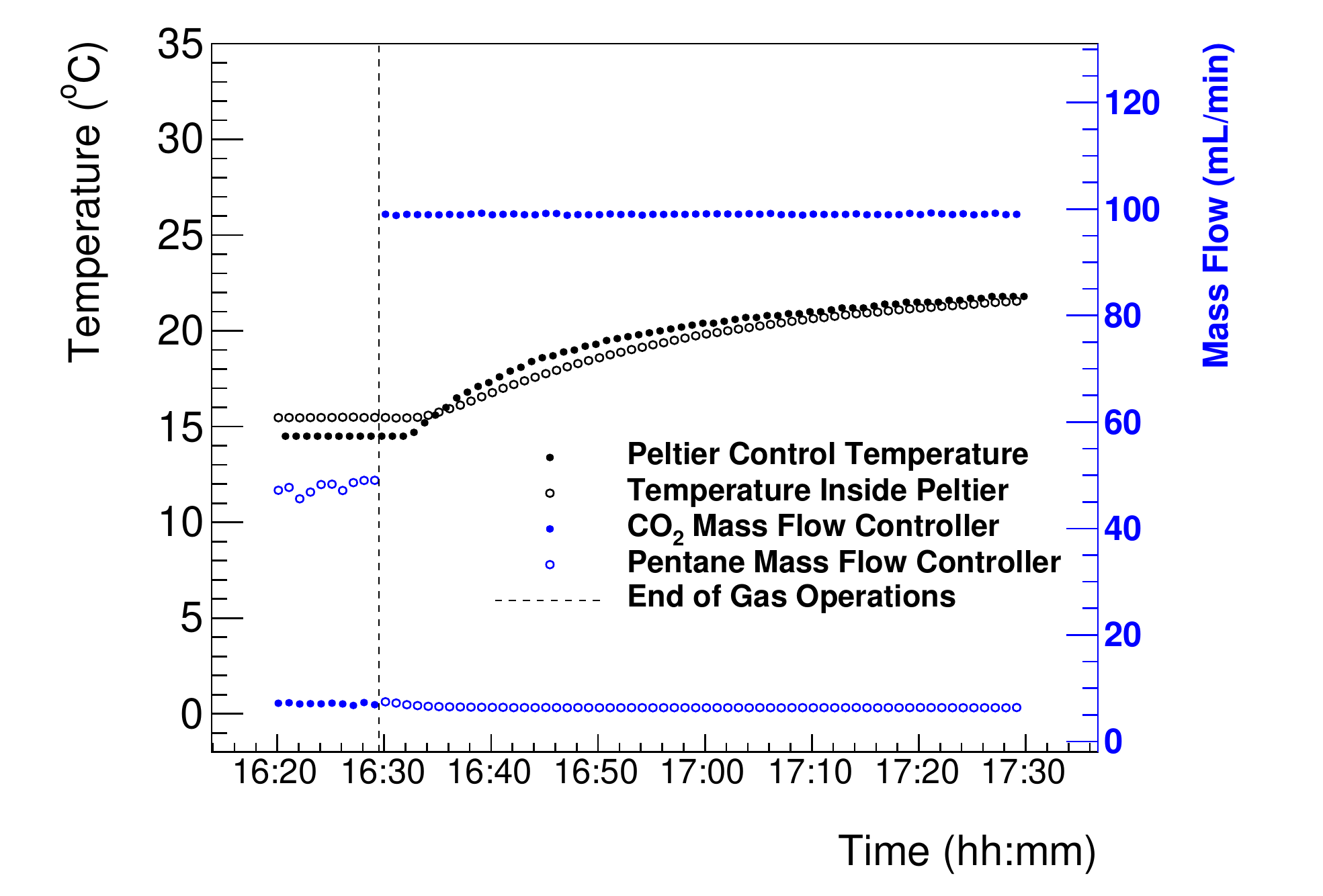}
                                \label{fig:StandardStop1}
    }%
    ~
    \subfigure[]{
        \centering
        \includegraphics[width=0.5\textwidth]{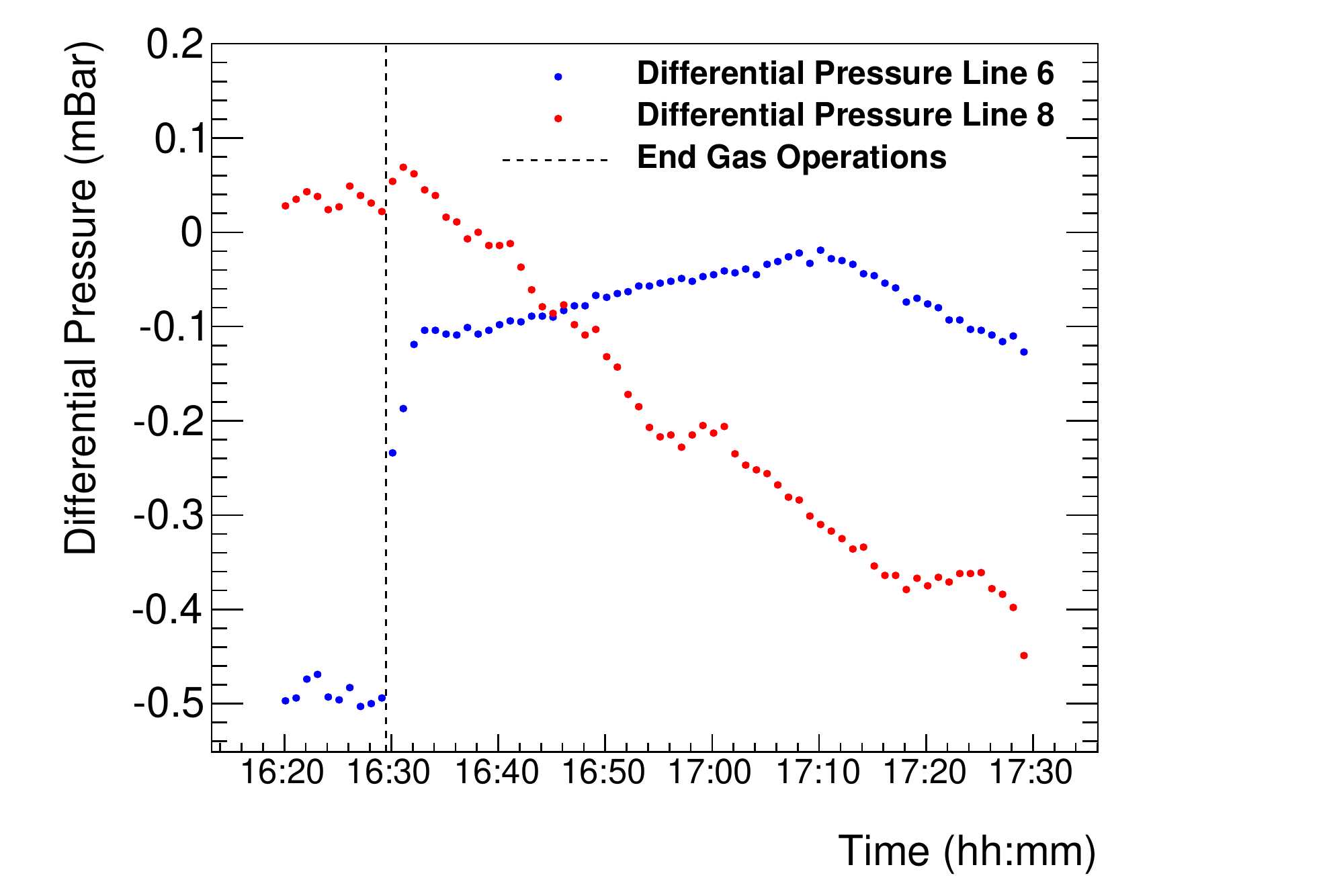}
                                \label{fig:StandardStop2}
    } 
               \caption{Temperatures and flow rates (a), and
                 differential pressures (b) during a standard start
                 sequence of the gas system. Temperatures and flow
                 rates (c), and differential pressures (d) during a
                 standard stop sequence of the gas
                 system.}\label{fig:gasrun}
\end{figure}

The normal start and stop sequences of a gas run using a single line (line 6) are 
illustrated in Figure~\ref{fig:gasrun}. From the \emph{Dormant} state, the
lines in operation are initially in a \emph{\co{} Flush} state with a flow rate of
approximately 100~\ml{}/min; the dashed line indicates the time at
which the \co{} gas in line 6 is replaced with the
\co{}:\np{}~mixture at an input \co{} flow rate of 45~\ml{}/min, as shown in
Figure~\ref{fig:StandardStart1}. The
Peltier set temperature settles at 14.5\deg{} while the
thermocouple readout is slightly higher (about 15\deg{} due to the
temperature gradient inside the gas pipe).  The differential pressure
of line 6 shows a sudden shift when switching from \co{} to
\co{}:\np{}-mixture, as displayed in Figure~\ref{fig:StandardStart2}. Line 8 is shown to illustrate the pressure trend for an unused line as a comparison. Figure~\ref{fig:StandardStop1}~and~\ref{fig:StandardStop2} show
reactions to a standard stop sequence. The dashed line indicates the
end of the gas run. Line 6 is returned to \emph{\co{} Flush} state and the
Peltier condenser system is stopped.

\begin{figure}
	\center
	\includegraphics[width=10cm]{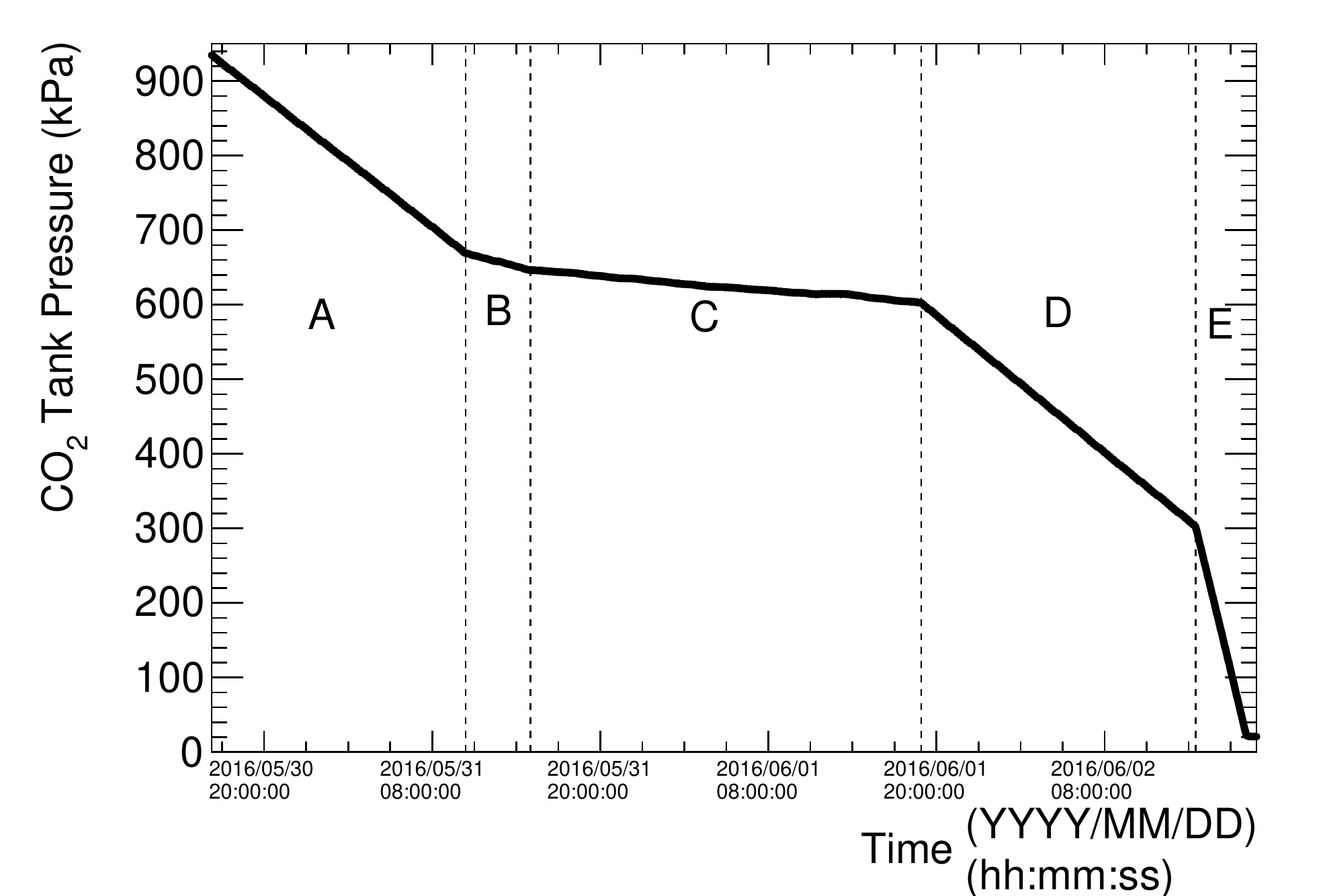}
	\caption{A typical plot observed from the monitoring tool as the \co{} tank empties. Each slope corresponds to a certain input flow rate (from left to right: \co{} flush of one line at 100~\ml{}/min (A), \np{}-\co{} flush of one line at 45~\ml{}/min (B), \np{}-\co{} flush of one line at 15~\ml{}/min (C), \co{} flush of one line at 100~\ml{}/min (D) and \co{} flush of five lines at 100~\ml{}/min each (E)).}
	\label{fig:EmptyCO2Tank}
\end{figure}

Based on our operational experience, a 2500~L \co{} tank is expected to last for approximately
2 months when operating a single gas line. During this time, the tank pressure remains constant as the pressurised \co{} is in liquid form.
As the tank is emptied, the liquid \co{} is depleted and the volume of the tank is filled with only gaseous \co{}. 
The pressure of the gaseous \co{} then decreases linearly, which is a signal that the tank is
almost empty, as shown in Figure~\ref{fig:EmptyCO2Tank}. 
Section A shows a \emph{\co{} Flush} at 100~\ml{}/min on 
the gas line connected to the \tgc{} detector. Section B shows 
the same gas line going to the \emph{Gas Operation} state (\co{}:\np{} mixture) at an input flow rate
of 45~\ml{}/min, then at 15~\ml{}/min in section C.
In section D, the line enters the \emph{\co{} Flush} state
once more at 100~\ml{}/min.
Finally, in section E, \emph{\co{} Flush} is initiated on five gas lines
simultaneously at 100~\ml{}/min each until the tank is completely empty. 
Monitoring
of the \co{} tank pressure therefore provides advance notice that a tank switch is needed,
helping to ensure uninterrupted gas-system operation.
The slow-control system sends an alert message when the pressure of
the \co{} tank goes below 3500~kPa so that the operator can anticipate when to stop the gas run and switch tanks before resuming operations.

\begin{figure}
	\center
	\includegraphics[width=10cm]{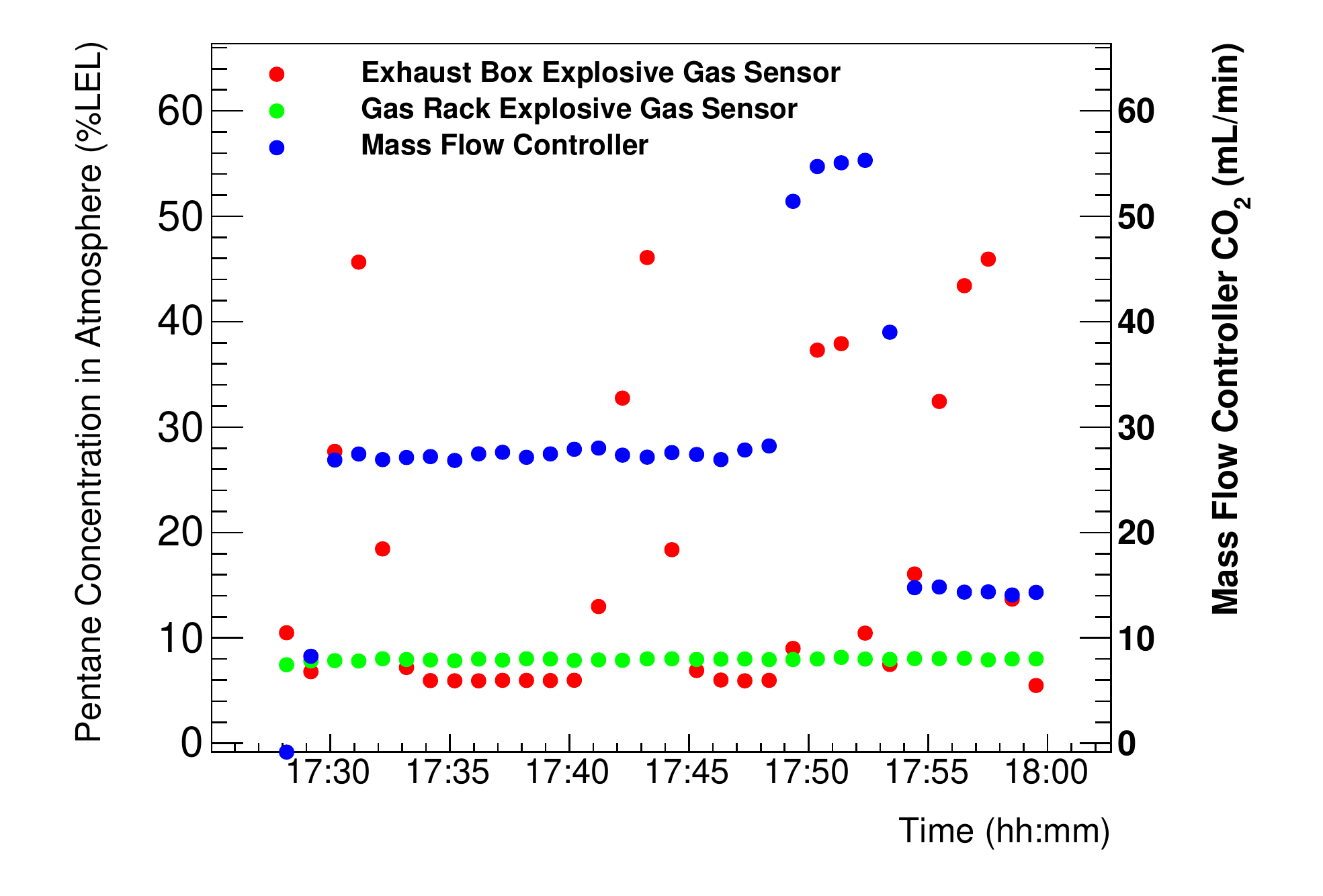}
	\caption{Tests of the explosive gas sensors with various MFC flow rates. The negative exhaust pressure is blocked 4 times, which creates peaks in the explosive gas sensor readout located in the exhaust box (red points), and gets released shortly after. The explosive gas sensor located in the gas rack (green points) does not react. Trials are carried out at different flow rates (blue points). As expected, the time for the concentration to build up is linearly correlated to the gas flow rate.}
	\label{fig:MFC_iTrans_test}
\end{figure}

Operational safety is critical in the presence of \np{} in the
laboratory; therefore, a prompt response of the explosive gas sensors is
needed. The response was tested in a controlled test where the dilution
box exhaust was blocked while \co{}:\np{} mixture was
flowing. Figure~\ref{fig:MFC_iTrans_test} shows the response of the
explosive gas sensors, one positioned for the purpose of this test inside the exhaust box, 
the other located at the bottom of the gas rack. As \co{} flows through the mixing apparatus 
(initially at a rate of 100~\ml{}/min, and then at 200~\ml{}/min and 50~\ml{}/min for short periods of time), 
the negative exhaust pressure is blocked in four instances such that the \co{}:\np{} mixture accumulates inside the exhaust 
box and is detected by the explosive gas sensors. The negative exhaust pressure is reinstated just before the 
lower explosive limit concentration (\%LEL) readout reaches 50\% to avoid any risks of explosion, which takes about two minutes (at 100~\ml{}/min). 
In all four instances, the explosive gas sensors are quick to respond to the presence of \np{} while the negative-pressure 
exhaust is efficient at evacuating the accumulated \np{}. 

\begin{figure}
    \centering
    \subfigure[]{
        \centering
        \includegraphics[width=0.5\textwidth]{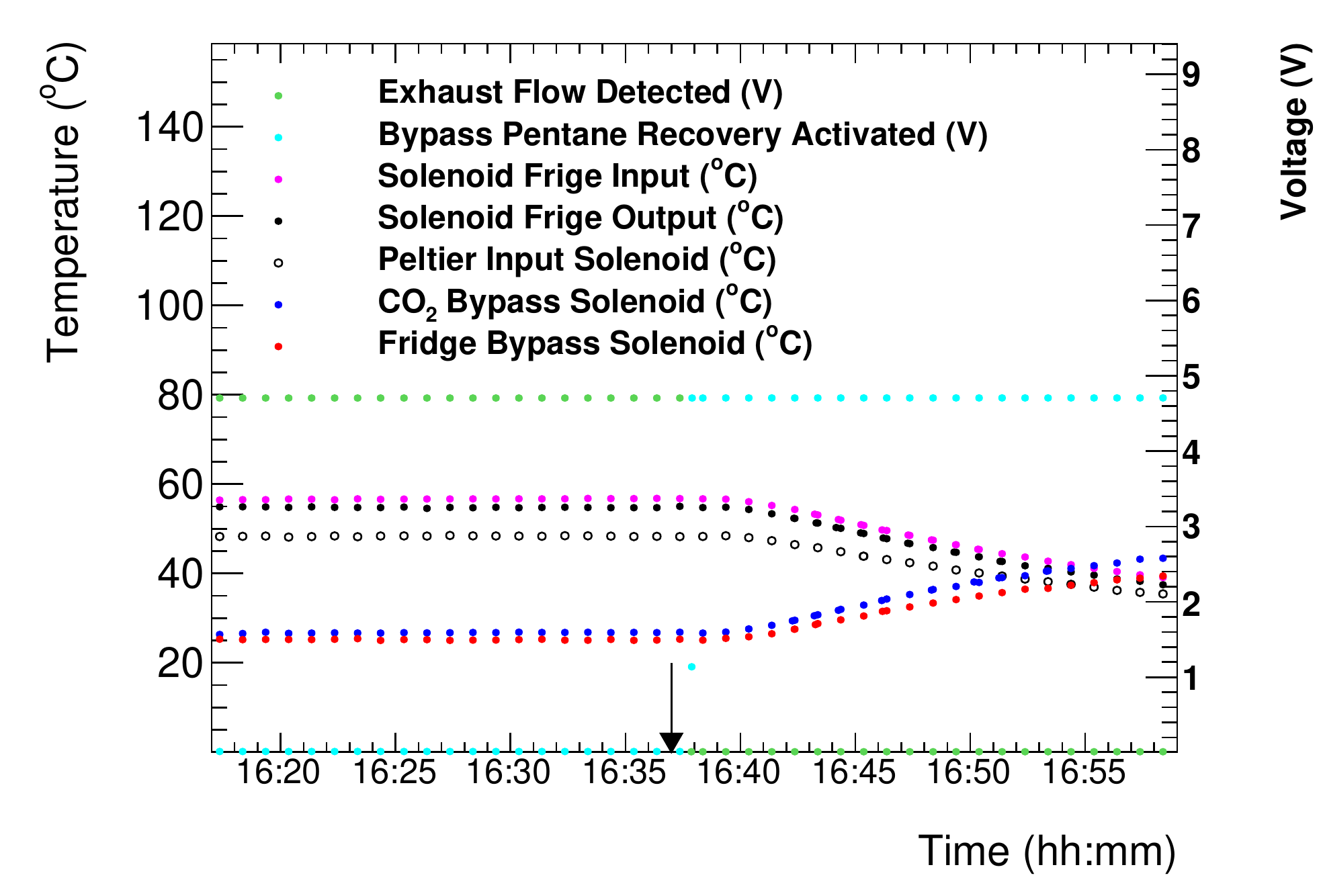}
				\label{fig:BlockageSimul1}
    }%
    ~ 
    \subfigure[]{
        \centering
        \includegraphics[width=0.5\textwidth]{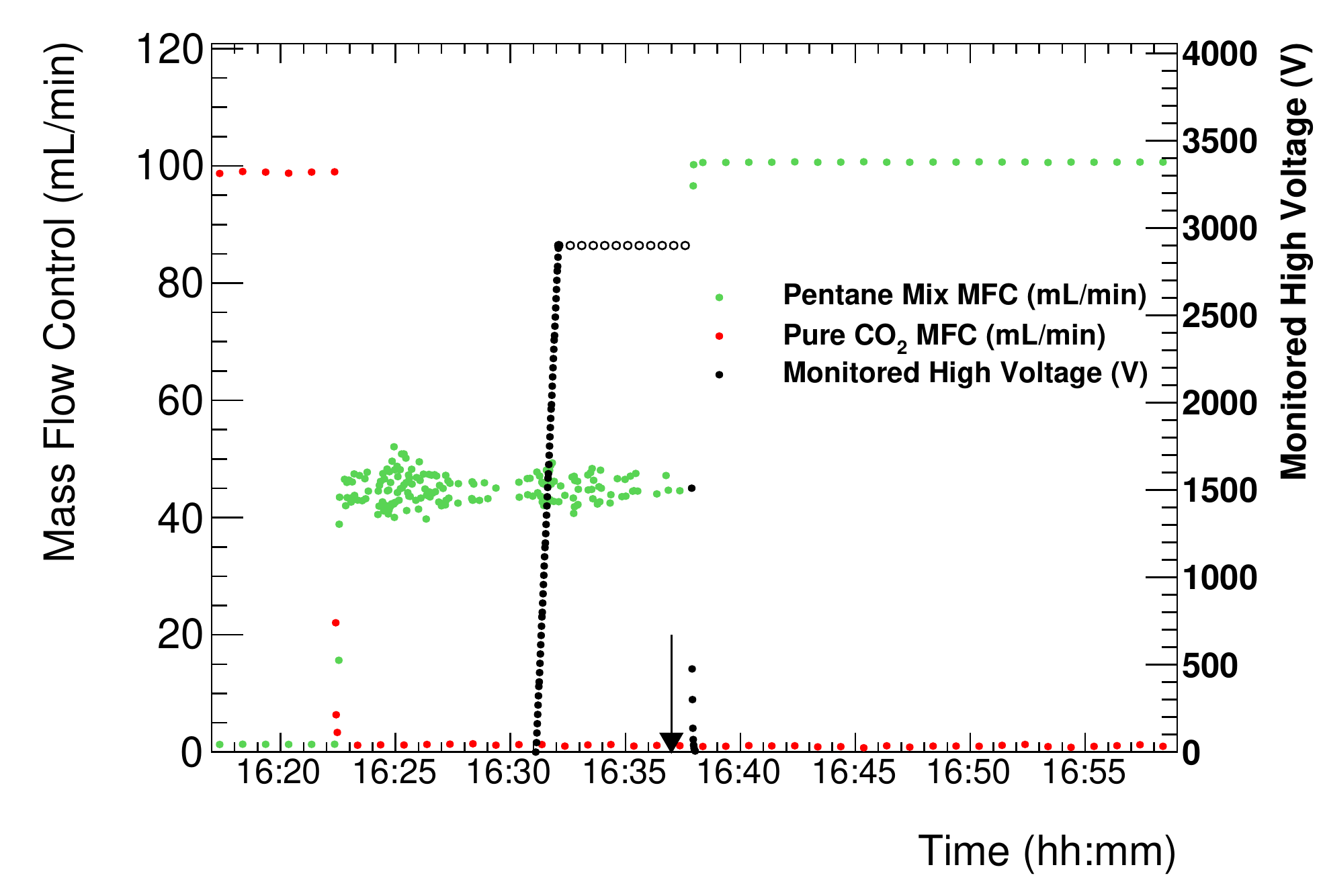}
				\label{fig:BlockageSimul2}
    }
		\caption{Response of the gas system to a simulated
                  exhaust blockage (indicated by an arrow). (a) Bypass solenoid valve system
                  response to a blockage event. Solenoid valves are
                  normally closed, so their temperatures increase when
                  they stay open. (b) High Voltage and MFC response to
                  a blockage event. Note that after the blockage event, the gas flowing in the system is pure \co{}.}
\end{figure}

As listed in Table~\ref{tab:StateMachineErrors}, there are three emergency cases in which the
slow-control system would 
put the gas system into a \emph{\co{} Bypass} state (excluding the timeout of a warning level error):
a high \np{} concentration detected by the explosive gas sensors, a solenoid valve overheat, or a lack of exhaust flow.
In the event of an exhaust failure during a run (whether due to a blockage, 
or a fan malfunction), the slow-control system would react immediately 
and bypass the \np{} mixer and recovery refrigerator, and turn 
off the high and low voltage supply. An exhaust failure was simulated for a single flowing line by manually blocking
the exhaust. In Figure~\ref{fig:BlockageSimul1}, where the time of the simulated blockage is indicated by the arrow, the temperature of
a solenoid valve can be used to tell whether it is open (high
temperature) or closed (low temperature). In Figure~\ref{fig:BlockageSimul2}, the high voltage
is turned off as soon as the exhaust blockage is detected by the flow sensor.
The mixing apparatus is bypassed and the MFC, which normally feeds the \co{}:\np{} mixture lines, starts flowing
pure \co{} at a rate of 100~\ml{}/min (the typical flow rate for the \co{} MFC) in each gas line as needed to flush the system.
These actions would vent a small amount of \np{} into the room, but would protect
the \tgc{} modules without creating a hazardous situation in the laboratory.

In the event of a high \np{} concentration detected by the explosive gas sensors, the same
action is taken as described above for the case of interrupted exhaust flow.  The quantity of \np{}
in the system at any given time, excluding the reservoir and recovery vessels, would be insufficient
to create a dangerous situation if it were all gradually vented into the laboratory.  In case of a
leak, flushing with \co{} protects the \tgc{} modules and helps to prevent the dangerous accumulation of \np{}.
The implementation of safety actions is fully automatic to ensure continuous operation of the facility.

\section{Conclusions}

The Canadian \tgc{} testing facility is being commissioned to use a cosmic-ray
hodoscope to characterise new muon detectors for the ATLAS Run~3 data-taking period planned to begin in 2021. A gas system was designed to supply the Canadian \tgc{}s with
the proper gas mixture needed for flushing and safe operations while in the testing facility. The mixing
apparatus creates a saturated \co{}:\np{} mixture at room temperature that 
is subsequently cooled in a Peltier TEC condenser system yielding the desired
mixture volume fraction of 45~\vp{} of \np{}. This innovative mixture apparatus design 
is simple, cost effective, and easy to build, providing reliable operation that permits swapping liquid \np{} reservoirs
without interrupting an ongoing run. The mixture fraction and performance 
of the system were characterised using two distinct methods: a mass measurement and a
gas chromatography analysis. This simple design could be used for upcoming gas detector testing facilities, such
as the one needed for the reception testing of the Canadian \tgc{}s at CERN.

Slow-control system software was written using \labview{} to ensure continuous, controlled
and reliable operation of the gas and HV/LV systems during cosmic-ray muon data
taking. The system software employs a state-machine framework that continuously monitors, 
characterises, and guides the operation of the system and the lab. The system 
takes as inputs various sensor readings, both from the apparatus itself, such 
as flow rates and pressures, and from the ambient conditions in the lab,
such as explosive gas readings, and manipulates various aspects of the system, 
such as high voltages and flow rates. The system software is designed to shut down safely with 
built-in hardware redundancy for failures involving \np{} gas risks. 
The slow-control system provides remote monitoring through a web interface
and alerts operators of warnings and errors via email and SMS. The Canadian \tgc{} 
testing facility is expected to commence operations in 2017.





\acknowledgments

The authors would like to thank members of the Tel Aviv University ATLAS sTGC group, especially Yan Benhammou and Meny Ben Moshe, and also members of the Weizmann Institute ATLAS sTGC group, especially Meir Shoa, Giora Mikenberg and Vladimir Smakhtin, for valuable discussions in the development of the McGill facility. Special thanks to the McGill Chemical Engineering laboratory for the gas chromatography instruments and technical expertise from Ranjan Roy and Andrew Golsztajn.
We gratefully acknowledge support from CFI, CRC, FRQNT, and NSERC (Canada).


\end{document}